%% file: 0_main.tex
\documentclass[journal,twoside,web]{ieeetran}
\usepackage{generic}
\usepackage{cite}
\usepackage{amsmath,amssymb,amsfonts}
\usepackage{algorithmic}
\usepackage{graphicx}
\usepackage{textcomp}
\usepackage{multirow}
\usepackage{array}
\usepackage{tabularx}
\usepackage{longtable}
\usepackage{adjustbox}
\usepackage{longtable}
\usepackage{tikz}
\usepackage[switch]{lineno}
\usepackage{url}

\def\BibTeX{{\rm B\kern-.05em{\sc i\kern-.025em b}\kern-.08em
    T\kern-.1667em\lower.7ex\hbox{E}\kern-.125emX}}
    
\newcolumntype{L}[1]{>{\raggedright\let\newline\\\arraybackslash\hspace{0pt}}p{#1}}
\newcolumntype{C}[1]{>{\centering\let\newline\\\arraybackslash\hspace{0pt}}p{#1}}
\newcolumntype{R}[1]{>{\raggedleft\let\newline\\\arraybackslash\hspace{0pt}}m{#1}}

\DeclareRobustCommand{\halfcheckmark}{\tikz\draw[scale=0.25,fill=black](0,.35) -- (.25,0) -- (1,.7) -- (.25,.15) -- cycle (0.75,0.2) -- (0.77,0.2)  -- (0.6,0.7) -- cycle;}

\begin{document}
{\renewcommand{\arraystretch}{1.5}

% Enable line numbers
% \linenumbers

\title{Securing Automated Insulin Delivery Systems: A Review of Security Threats and Protective Strategies
}
\author{Yuchen Niu and Siew-Kei Lam
% \IEEEmembership{Senior Member, IEEE}
% \thanks{This paragraph of the first footnote will contain the date on 
% which you submitted your paper for review. It will also contain support 
% information, including sponsor and financial support acknowledgment. For 
% example, ``This work was supported in part by the U.S. Department of 
% Commerce under Grant BS123456.'' }
% \thanks{The next few paragraphs should contain 
% the authors' current affiliations, including current address and e-mail. For 
% example, F. A. Author is with the National Institute of Standards and 
% Technology, Boulder, CO 80305 USA (e-mail: author@boulder.nist.gov). }
\thanks{Yuchen Niu and Siew-Kei Lam are with the College of Computing and Data Science, Nanyang Technological University, Singapore 639798, Singapore (e-mail: yuchen017@e.ntu.edu.sg (corresponding author); assklam@ntu.edu.sg).}

\thanks{This is the authors’ accepted manuscript of a paper accepted for publication in
Computers \& Security (Elsevier). The final version of record is available at:
https://doi.org/10.1016/j.cose.2025.104733}
}

\maketitle

\begin{abstract}
Automated Insulin Delivery (AID) systems represent a significant advancement in diabetes care and wearable physiological closed-loop control technologies, integrating continuous glucose monitoring, control algorithms, and insulin pumps to improve blood glucose level control and reduce the burden of patient self-management. However, their increasing dependence on wireless communication and automatic control introduces security risks that may compromise patient privacy or result in life-threatening treatment errors. This paper presents a comprehensive survey of the AID system security landscape, covering technical vulnerabilities, regulatory frameworks, and commercial security measures. In addition, we conduct a systematic review of attack vectors and defence mechanisms proposed in the literature, following the PRISMA framework. Our findings highlight critical gaps, including the lack of specific security evaluation frameworks, insufficient protections in real-world deployments, and the need for comprehensive, lightweight, and adaptive defence mechanisms. We further investigate available research resources and outline open research challenges and future directions to guide the development of more secure and reliable AID systems. By focusing on AID systems, this review offers a representative case study for examining and improving the cybersecurity of safety-critical medical wearable systems.

\end{abstract}


\section{Introduction}
\label{sec:introduction}
The management of Type 1 Diabetes (T1D), a chronic condition where pancreatic beta cells produce insufficient insulin for regulating blood glucose (BG) levels, remains a significant challenge for millions of individuals worldwide \cite{IDF-T1D}. As there is no realistic biological cure in the near future, the treatment primarily involves intensive insulin therapy to maintain BG levels within a safe range, thereby mitigating the risk of serious complications caused by hyperglycemia or hypoglycemia \cite{132}.  However, this regimen of frequent BG measurement and insulin adjustments is time-consuming and error-prone for patient self-management. To address these challenges, automated insulin delivery (AID) systems, also known as artificial pancreas, have emerged as a promising “technical” solution \cite{150}. 

AID system integrates a continuous glucose monitor (CGM), a control algorithm, and an insulin pump to automatically adjust insulin delivery based on real-time glucose measurements. With recent technological advancements, these systems have evolved into networked and intelligent medical wearables that improve glycemic outcomes and quality of life for individuals with diabetes \cite{129}\cite{Godoi2023Glucose}. Nevertheless, the increasing device complexity, coupled with their reliance on wireless connectivity and automated control, introduces significant safety and security concerns \cite{130}. Ensuring the trustworthiness and resilience of AID systems against a broad spectrum of threats is essential for safeguarding patient health and enabling wider clinical adoption.

Threats to the reliability of AID systems primarily arise from two sources: unintended failures and malicious attacks. Unintended failures result from system limitations \cite{150} and malfunctions \cite{151}. For example, physiological constraints on delayed insulin absorption and postprandial rapid glucose fluctuations can limit the effectiveness of insulin delivery. Technological and behavioural limitations, such as sensor noise, signal dropouts, miscalibration, or pressure-induced sensor attenuation (e.g., lying on the sensor during sleep), may lead to inaccurate glucose readings and inappropriate dosing. Additionally, malfunctions with CGM and insulin pumps (e.g., early sensor failure, occlusion of insulin infusion sets, and cartridge errors) can further compromise system performance.

Malicious attacks, on the other hand, involve adversaries exploiting vulnerabilities in AID systems to intercept sensitive patient data or manipulate insulin delivery. These risks have escalated with the adoption of wireless communication in AID systems \cite{23}. Demonstrated attacks include unauthorised remote control \cite{143} and false CGM reading injection \cite{158}. Early experiments have shown the feasibility of intercepting communication and launching lethal insulin doses through unencrypted wireless channels at least 20 meters away \cite{1}\cite{100}. Recognising these risks, the U.S. Food \& Drug Administration (FDA) issued warnings in 2019 regarding cybersecurity weaknesses in AID systems \cite{102}, and several commercial products have been recalled due to security concerns \cite{101}.

\input{Table_1-1}

Strengthening the reliability of AID systems requires a thorough understanding of existing and emerging threats, a critical evaluation of defence mechanisms, and the identification of gaps that must be addressed. Review papers on the reliability of AID systems published in the past decade, as summarized in Table \ref{tab:table1-1}, most primarily focused on discussing unintended failures  \cite{91} \cite{89} \cite{153} or discussing threats from malicious attacks without analysing specific attack vectors and defence mechanisms  \cite{150}\cite{29} \cite{88}. While several studies provided general overviews of cyber vulnerabilities and potential countermeasures, they lack in-depth examinations of related studies \cite{15} \cite{92} \cite{87}\cite{154}. Nazzal et al. \cite{152} conducted a thorough literature review of software verification methods for AID systems; however, broader security defence strategies remain unexplored. Besides, while some existing reviews on Internet of Medical Things (IoMT) security briefly discuss AID systems, they do not adequately address their distinctive risks \cite{82}\cite{93}\cite{94}. 

Compared to most other safety-critical IoMT systems, AID systems present unique security challenges. First, the physiological closed-loop control introduces substantial interpatient variability and unpredictable external factors, complicating threat detection and leaving a narrow window for preventing attacks from impacting drug delivery. Second, AID wearables operate under stringent power, computation, and memory constrained, limiting the feasibility of resource-intensive security strategies. These challenges not only amplify the impact of potential threats but also complicate the development of effective defence mechanisms. Notably, no prior reviews comprehensively examine security threats and protective strategies for wearable physiologic closed-loop control systems. 

To address this gap, this study systematically reviews the security of AID systems, providing an in-depth analysis of attack vectors, existing defence strategies, and open research challenges. We introduce a taxonomy that categorises security threats and defence mechanisms in AID systems, as illustrated in Figure \ref{Taxonomy}. With the growing adoption of AID technologies and the expanding body of related research, this review not only advances understanding of AID security but also serves as a representative case study for strengthening the cybersecurity of other medical wearable systems.

The remaining paper is organised as follows (see Figure \ref{graphic overview}). Section \ref{sec:method} presents the methodology and materials used in the review. Section \ref{sec:background} examines the security status of AID systems from technical, legal, and commercial perspectives, establishing a foundational understanding of the field. Section \ref{sec:attack} reviews existing and emerging attack vectors, and Section \ref{sec:defence} analyses defence strategies proposed in the literature. Section \ref{sec:resources} discusses current evaluation methods and available research resources. Section \ref{sec:key_findings} discusses key findings and research gaps, and Section \ref{sec:challenges} outlines future directions for strengthening the security of AID systems. 

\begin{figure*}[t]
    \centering
    \includegraphics[width=0.8\textwidth]{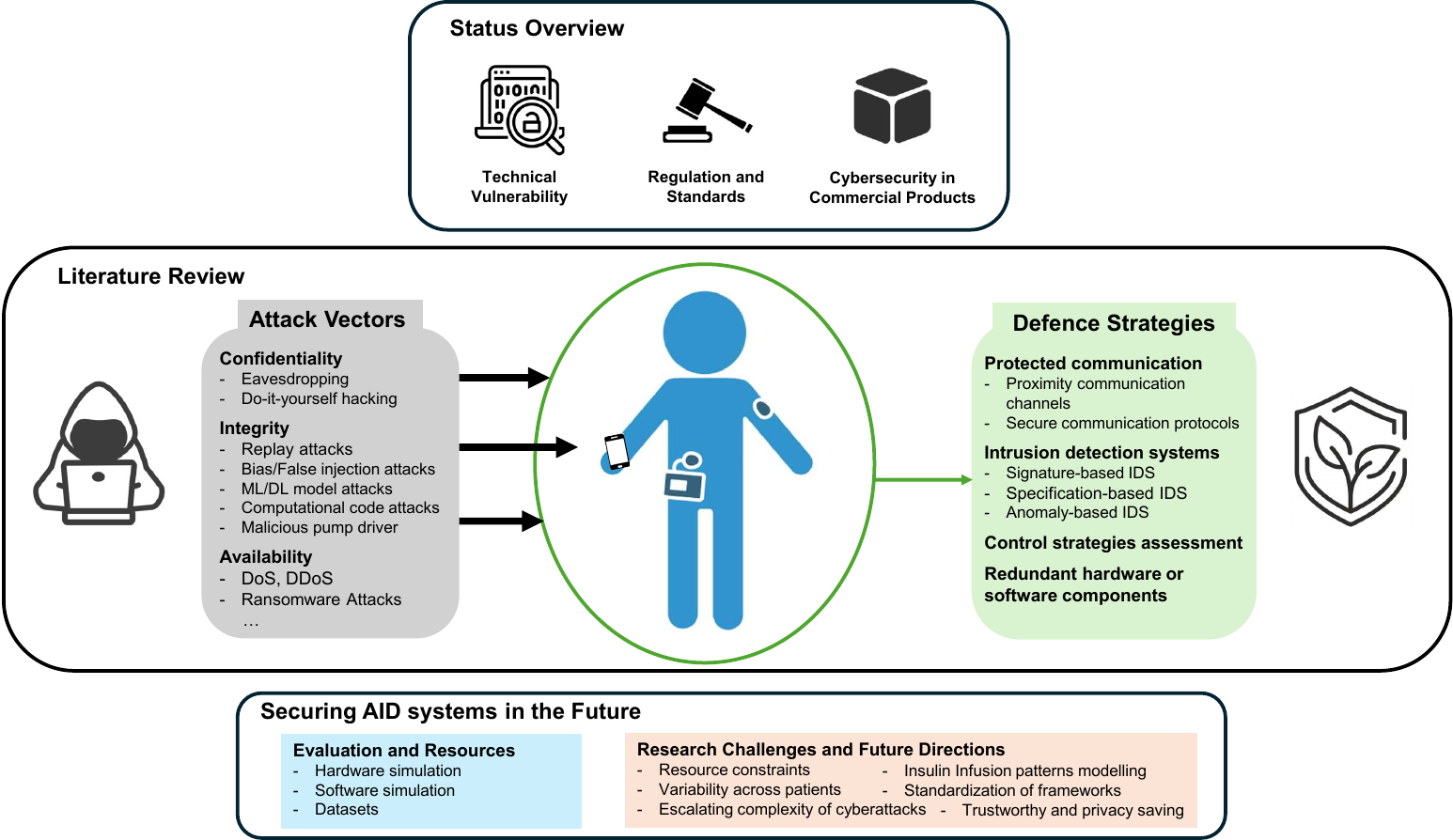}
    \caption{Overview of the paper.  }
\label{graphic overview}
\end{figure*}

\begin{figure*}[t]
    \centering
        \makebox[\textwidth][c]{%
        \includegraphics[width=0.9\linewidth]{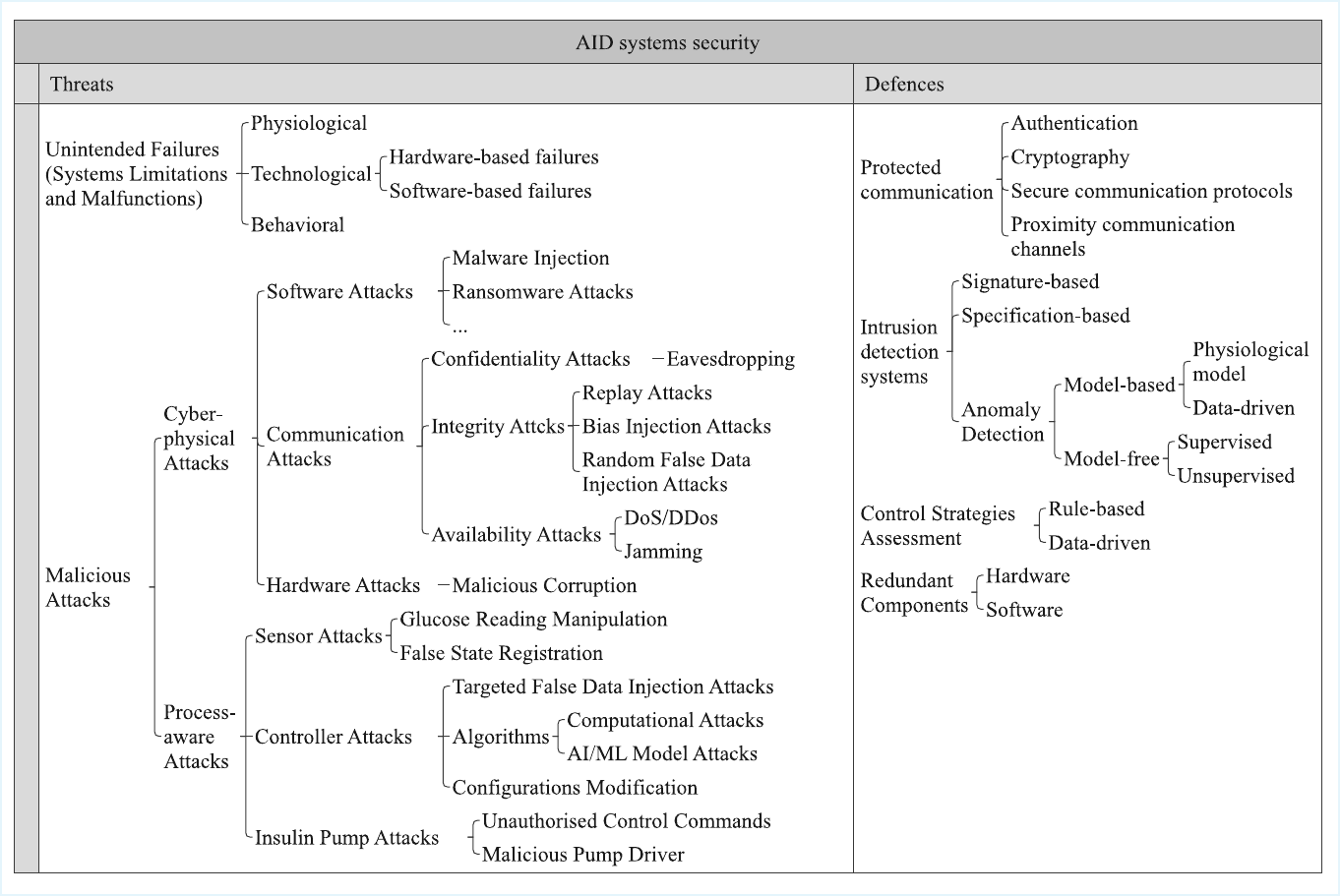}
    }
    \caption{Taxonomy of security threats and defence strategies in AID systems.}
\label{Taxonomy}
\end{figure*}

\section{Methodology and Materials}
\label{sec:method}
To comprehensively assess the security landscape of AID systems, this paper adopts a hybrid methodology, integrating elements of both survey and systematic review.
The review begins with a broad investigation into AID systems cybersecurity, covering technical vulnerabilities, regulatory frameworks, and security measures in commercial products. 
The materials considered extend beyond peer-reviewed journals and conference papers to include diverse sources such as clinical studies, industry reports, regulatory standards, product manuals, and relevant online resources.
Following this, the PRISMA framework \cite{156} was employed for a systematic review of attack vectors and defence mechanisms proposed in the literature for AID systems security. To guide this analysis, the following research questions (RQs) were formulated:

\begin{itemize}
    \item RQ1: What potential attack vectors and risks do AID systems face?
    \item RQ2: How do existing defence mechanisms address these risks?
    \item RQ3: How practical and robust are these solutions in protecting AID systems from cybersecurity threats?
\end{itemize}

A systematic literature search was conducted across multiple databases, including Scopus, Google Scholar, Web of Science, IEEE Xplore, and PubMed. Boolean operators and keyword variations were applied, as summarized in Table \ref{tab:table1-2}. The review focused on publications from 2010 to the collection date (February 11, 2025), with studies outside this range filtered out during the initial search.

\input{Table_1-2}

All retrieved papers underwent a screening process (PRISMA diagram in Figure \ref{PRISMA}) based on predefined inclusion and exclusion criteria.

Inclusion criteria:
\begin{itemize}
    \item Peer-reviewed journal articles, conference papers, or book chapters;
    \item Published between 2010 and 2025;
    \item Studies proposing attack vectors, defence mechanisms, or risk evaluations for AID systems
\end{itemize}

Exclusion criteria:
\begin{itemize}
    \item Papers not written in English
    \item Review papers;
    \item Irrelevant articles;
    \item Duplicate articles
\end{itemize}

\begin{figure}[t]
    \centering
    \includegraphics[width=1\linewidth]{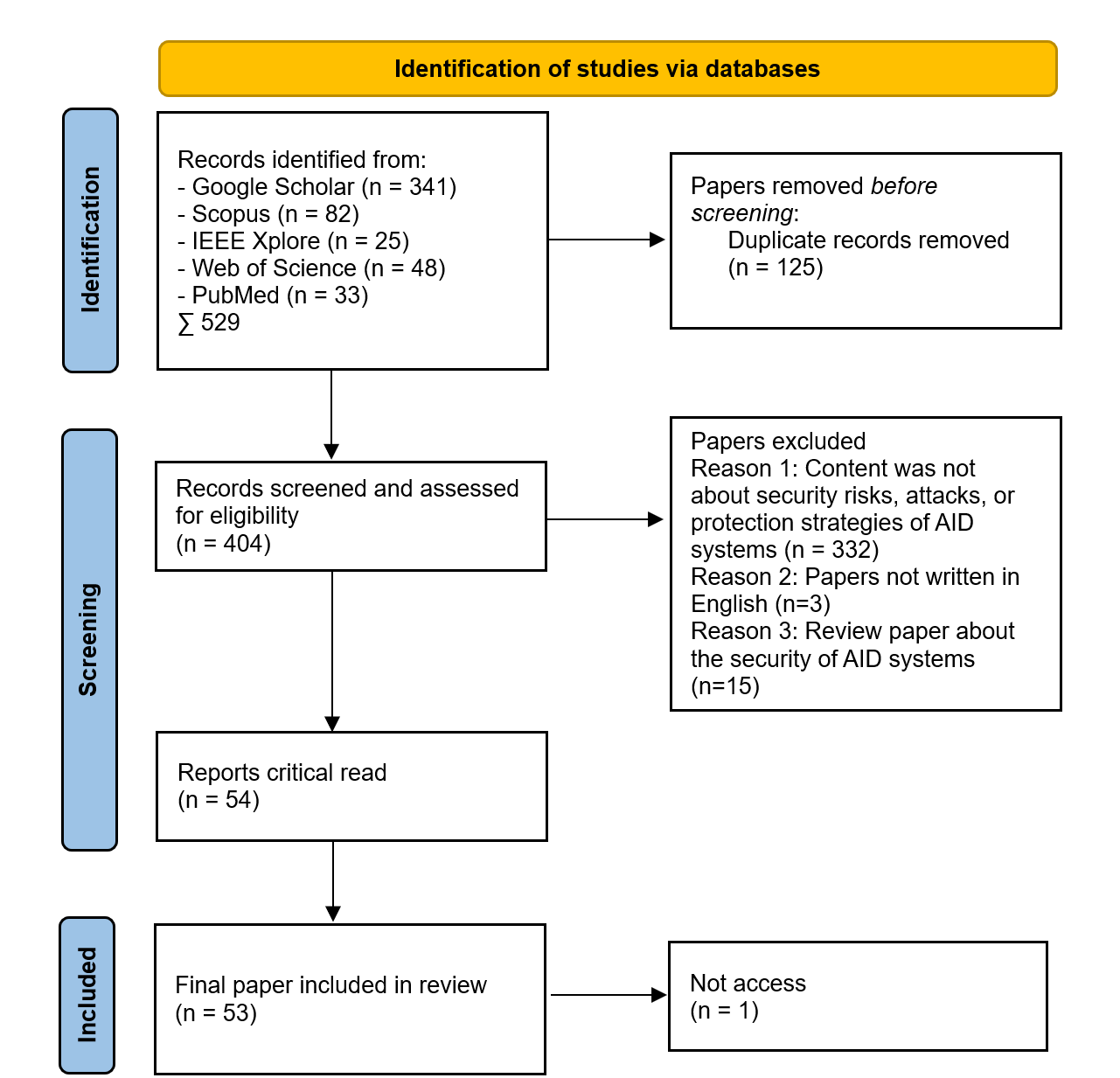}
    \caption{PRISMA diagram for the screening process.}
\label{PRISMA}
\end{figure}

From the initial database search, 53 relevant papers were identified. To ensure comprehensive coverage, backward and forward citation tracking was conducted, yielding 23 additional studies. These were initially missed due to: (1) Broad scope—studies focusing on medical devices, IoT-based healthcare systems, or cyber-physical systems that take AID systems as a case study; and (2) Terminology differences—studies referring to AID systems under alternative terms such as wireless insulin pump systems or sensor-augmented pumps.

Finally, 76 papers met the inclusion criteria for analysing attack vectors and defence mechanisms in AID systems. The publication trends over the past 15 years are visualized in Figure \ref{paper_publication}. Notably, a significant rise in proposed defence mechanisms has been observed in the past five years.

\begin{figure}
    \centering
    \includegraphics[width=1\linewidth]{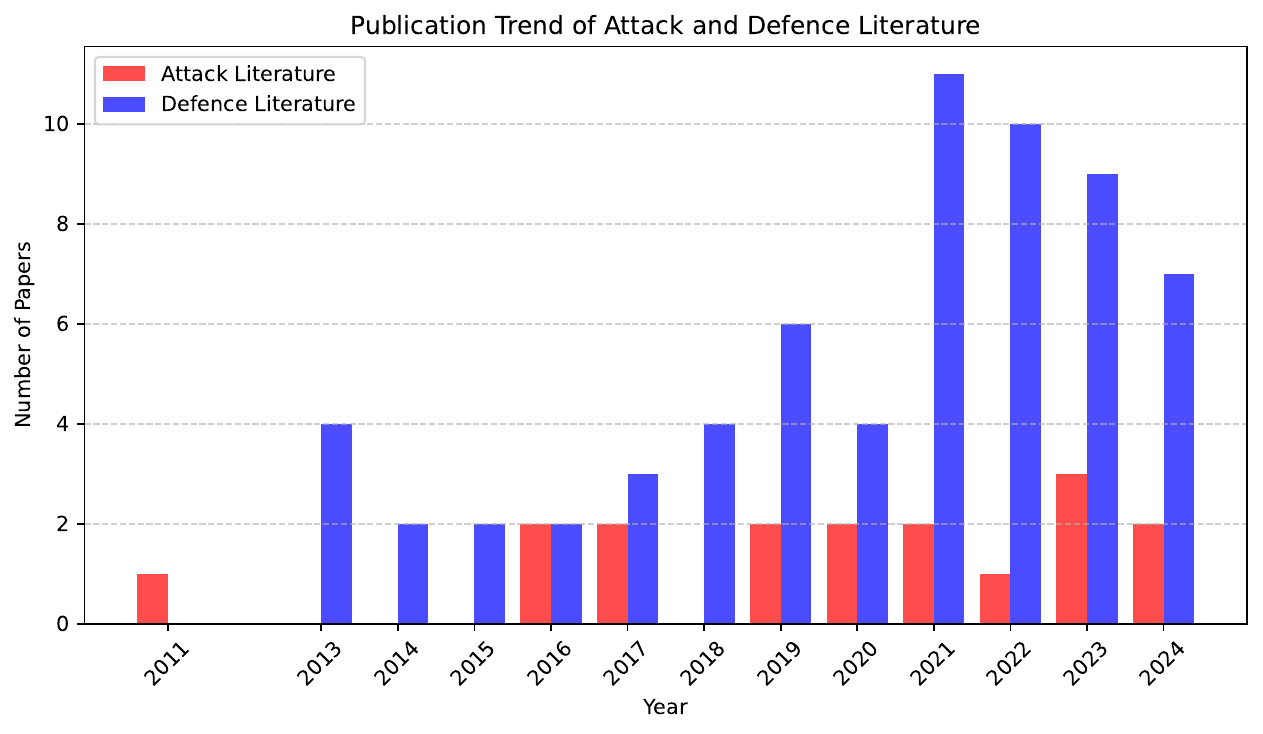}
    \caption{Publication trend of attack and defence literature.}
\label{paper_publication}
\end{figure}

\section{Background}
\label{sec:background}
\subsection{Technical Description and Vulnerabilities}

\begin{figure}[t]
    \centering
    \includegraphics[width=0.5\textwidth]{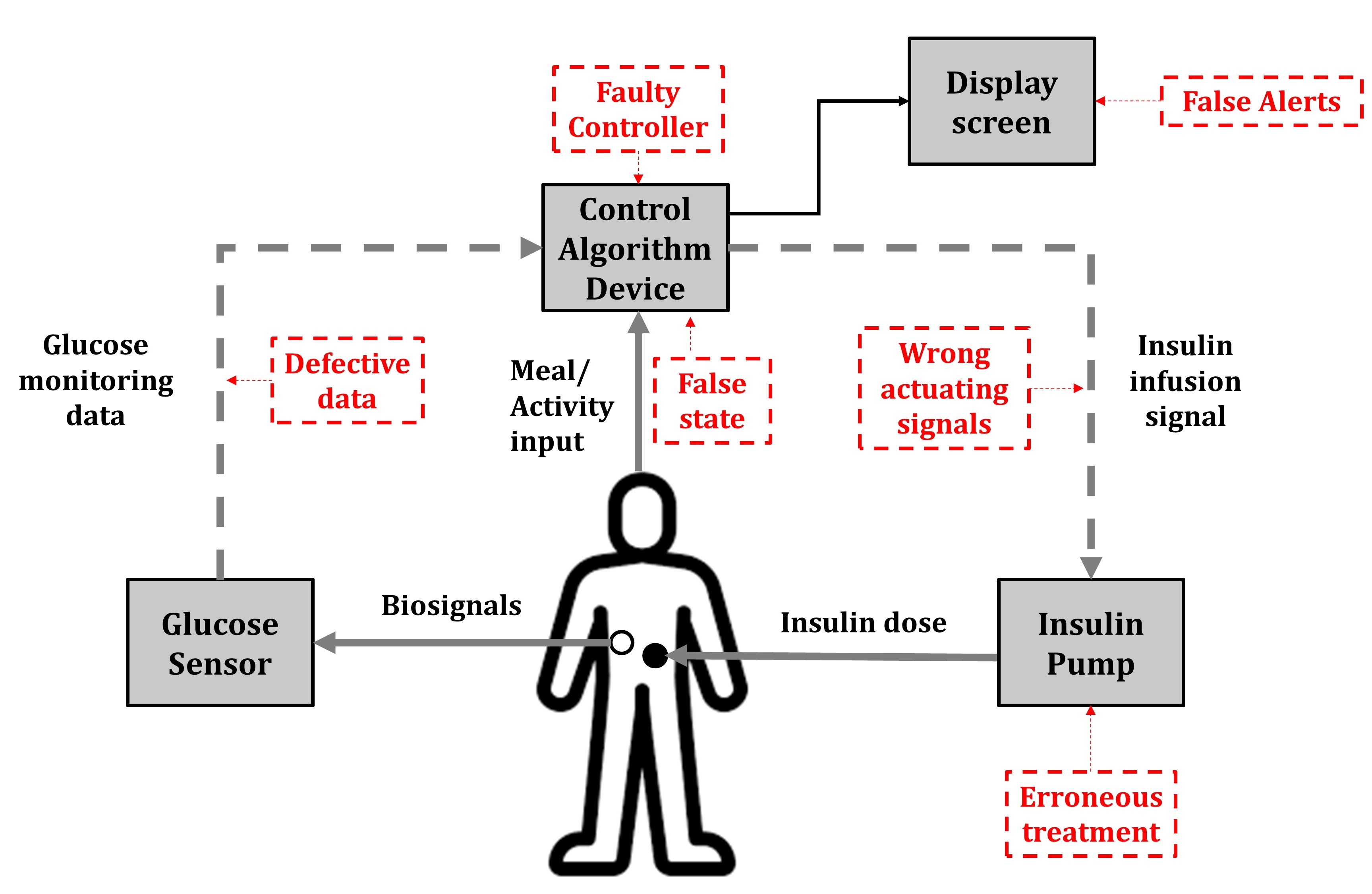}
    \caption{An overview of a hybrid closed-loop insulin delivery
system, illustrating core components, workflow, and potential functional vulnerabilities. Dotted lines indicate wireless communication links between modules, while solid lines represent direct attachments to the patient’s body or integrated elements - adapted from \cite{80}.}
\label{overview}
\end{figure}

AID combines a CGM, an insulin pump, and a controller algorithm to calculate and deliver suitable insulin doses, helping to maintain blood glucose levels within the target range. Figure \ref{overview} shows the workflow of a hybrid AID system, the paradigm common among current available products. In this system, the CGM periodically (e.g., every 5 minutes) transmits the measured glucose levels to the control algorithm device, which calculates basal insulin and corrective boluses in response to sensor glucose values and meal registration, and then directs the subcutaneous insulin pump to deliver the specified dose into the body \cite{59}. Unlike full AID systems that automate all insulin delivery, hybrid systems require the user to administer prandial insulin before or at the time of meals to control post-meal blood glucose spikes \cite{11}.  Full AID systems, though promising, still face challenges in managing postprandial hyperglycemia and exercise-related glycemic fluctuations \cite{14}.  Despite different automation levels, both hybrid and full AID systems share core modules and rely on wireless connectivity and software control, exposing them to similar cybersecurity vulnerabilities.

Cybersecurity concerns the protection of digital assets. In the context of AID systems, these assets include physiological and treatment data transmitted wirelessly between system components, the software controlling processes, and patient information stored on the device. Vulnerabilities may arise in both wireless communication channels and individual system modules, as detailed in the following subsections.

\subsubsection{Wireless communication}
Communication among the three core components of AID systems primarily relies on wireless transmissions for wearable convenience, making it possible for attackers to intercept, disrupt, or manipulate signals from a distance. Secure information exchange requires communication protocols that provide robust encryption, authentication, and resilience under emergencies. However, as examined in \cite{77}, protocols commonly applied in AID systems are often lightweight but fail to meet these security requirements. For example, the widely used Bluetooth Low Energy (BLE) protocol is vulnerable to replay and man-in-the-middle attacks, allowing adversaries to eavesdrop or inject faulty glucose readings and control commands~\cite{137}.

\subsubsection{CGM}
A CGM system continuously tracks BG levels in the patient's body and transmits the collected physiological data to the controller every few minutes. It consists of a disposable glucose sensor, placed under the skin to measure glucose levels, and a transmitter attached to the sensor that sends the data to the control algorithm platform. Accurate BG readings are crucial for safe insulin delivery, as any incorrect values can potentially result in harmful dosing decisions. However, a compromised CGM system may operate maliciously by providing inaccurate BG levels, either due to device tampering or data manipulation during wireless transmission \cite{1}\cite{100}\cite{59}.

\subsubsection{Control Algorithms}
The controller computes the recommended insulin doses and transmits infusion commands to the insulin pump.  It incorporates patient physiological models to estimate glucose-insulin dynamics and control algorithms to determine optimal insulin doses. These computations rely on real-time sensor data, historical records, and estimated patient status (e.g., eventual BG, Insulin on Board (IOB)). Control algorithms may be deployed on smartphones, tablets, personal computers, or integrated into insulin pumps, which are often the platforms for user monitoring device activities. Three main security threats are associated with the controller: 
\begin{itemize}
  \item Data Manipulation: The stored records, received sensor data, or meal and activities inputs could be tampered with or falsely registered, compromising the accuracy of the controller’s calculations \cite{42}\cite{2}. Additionally, alerts on the display screen could be maliciously triggered or disabled, prompting users to turn off or ignore the security monitoring system, thus heightening system vulnerability. 
  \item Compromising controller algorithms: The core logic and configuration of control algorithms may be altered, resulting in incorrect insulin dosing \cite{4}.
  \item Platform Vulnerabilities: The device running the control algorithm could be compromised by malware, such as viruses or worms, which may interfere with the proper functioning of the AID systems \cite{128}.
\end{itemize}

\subsubsection{Subcutaneous Insulin Pump}
A subcutaneous insulin pump includes a reservoir, which stores insulin and supplies it to the delivery system, an infusion set consisting of a thin tube and a subcutaneous cannula for insulin delivery, and a microcontroller unit that orchestrates the pump’s functions by executing programmed commands based on input from the control platform. Security concerns associated with the pump arise from device malfunctions (e.g., infusion site failures)\cite{9}\cite{138} and malicious control I/O operations of the pump\cite{100}.

\subsection{Regulation Standards and Guidelines}

The regulatory landscape governing the cybersecurity of AID systems is evolving, with growing recognition of cybersecurity as a critical component of patient safety. This section examines the regulatory frameworks of the United States (US) and the European Union (EU). Given the global nature of medical device markets, many other countries have adopted regulations that closely align with these standards. Appendix \ref{app:regulation} summarises the core regulatory standards and guidelines relevant to AID system cybersecurity in these regions. 

In the US, the FDA oversees regulatory compliance for medical devices. Existing legislation and FDA guidance documents highlight cybersecurity as a key consideration in approval and inspection processes. Notably, the FDA has issued guidance for premarket submissions titled "Cybersecurity in Medical Devices: Quality System Considerations and Content of Premarket Submissions" \cite{food2023cybersecurity}, and for postmarket submissions titled "Postmarket Management of Cybersecurity in Medical Devices" \cite{FDA2016cybersecurity}. These documents provide prescriptive and structured approaches for manufacturers to conduct risk management and implement cybersecurity controls. In addition, the NIST's cybersecurity framework \cite{NIST-CSF} provides a structured catalogue of typical cybersecurity activities and controls that enterprises should consider.

In the EU, the Medical Devices Regulation (MDR) \cite{EU-MDR-2017} serves as the primary legislative framework, supplemented by other cybersecurity standards. Annexe I of the MDR, General Safety and Performance Requirements (GSPR), requires that medical devices must not fail due to security attacks and that necessary cybersecurity protections must be in place. To facilitate implementation, the Medical Device Coordination Group (MDCG) issued the guidance document “MDCG 201916: Guidance on Cybersecurity for medical devices” \cite{medical2019mdcg}, which provides recommendations to manufacturers on fulfilling MDR cybersecurity requirements. Compared with FDA guidance, MDCG offers relatively high-level principles on how to operationalise these activities \cite{skytterholm2025cybersecurity}.

While cybersecurity regulations in the US and EU both require manufacturers to identify, assess, and mitigate cybersecurity risks across the product lifecycle, responsibility for comprehensive threat modelling and protective measures implementation rests with the manufacturer. So the approach and depth are determined by the manufacturer and subject to regulatory review. This creates challenges in mapping the broad landscape of security requirements to specific devices and maintaining up-to-date threat models. Moreover, current guidelines lack formal mechanisms for establishing confidence in a product or system's capacity to defend against or recover from cyber attacks through independent assessment \cite{17}.

To address these limitations, independent organisations and researchers have proposed complementary approaches. For example, the nonprofit organisation Diabetes Technology Society (DTS) introduced the DTSec (DTS Cybersecurity Standard for Connected Diabetes Devices) to define AID-specific security requirements and enable independent compliance assessment \cite{DTSec-Standard}. In addition, to assess the cyber risk against the up-to-date threat landscape, Shen \cite{shen2025risk} proposed a cyber risk assessment method that evaluates Manufacturers Disclosure Statement for Medical Device Security (MDS2) data sheets against updated cyber threat intelligence using MITRE ATT\&CK. These initiatives highlight potential directions for evolving regulation toward more comprehensive and concrete cybersecurity compliance regulations.

\subsection{Commercial Products}

With the rising prevalence of diabetes and continued technological advancement, global demand for AID systems is rapidly increasing, with the market valued at US\$2.34 billion in 2023 and projected to reach US\$5.80 billion by 2032 \cite{98}. Nevertheless, safety and security challenges persist. For example, by 2023, 24 device failures and recalls were documented for a single AID system manufacturer  \cite{153}. A cross-sectional survey \cite{130} revealed that concerns regarding privacy and reliability are one of the major factors that hinder broader adoption of AID systems. Appendix \ref{app:products} summarises the core components, functions, and security measures of representative products. Although the cybersecurity capabilities of leading products have improved, they remain limited in terms of security protection, risk management, and monitoring. 

Experimental studies have demonstrated that two early-generation AID products lack standard cryptographic mechanisms, allowing full reverse-engineering of their wireless protocols and exposing them to eavesdropping, replay, and data manipulation attacks\cite{1}\cite{41}. These vulnerabilities resulted in official warnings of use \cite{142} and product recalls \cite{143}. Although more recent products employ encryption and authentication to protect wireless communication, the level of security depends on factors such as method strength, key management, and system integration. Even with strong encryption and authentication, vulnerabilities remain to advanced threats like side-channel attacks, denial-of-service (DoS) attacks, and physical tampering.

To address advanced or unknown threats that may bypass safeguards, robust risk management is essential for ensuring device reliability under attack. However, risk management functions in current products are limited. Insulin pumps incorporate preset minimum and maximum dosage and infusion rate limits, while systems like the Medtronic 780G suspend insulin delivery when glucose levels drop below a predefined threshold. Yet, these threshold-based approaches are ineffective against more sophisticated attacks. Research has demonstrated that personalised insulin dose manipulation attacks can subtly alter insulin delivery in ways that are harmful to specific patients at particular times while remaining within general safety constraints  \cite{10}. In addition, false data injection attacks can falsify glucose readings to suppress alarm activation, and computational attacks can modify algorithm logic \cite{100}. Detecting these attacks requires identifying abnormal patterns rather than comparing with predefined thresholds.

Meanwhile, the main responsibility for monitoring and securing devices is explicitly assigned to patients themselves as indicated in user manuals \cite{CamDiab2025}.  However, this expectation is often unrealistic, especially during sleep or other daily activities. Besides, the absence of real-time displays in CGM systems further limits patients’ ability to detect discrepancies between glucose readings and data relayed to the control platform.  

Given these limitations, there is a pressing need for manufacturers to establish sufficient security measures to prevent, detect, and respond to security threats in real time without relying solely on patient intervention. Strengthening security in this manner is essential to improving system reliability and fostering user trust.

\section{Attack Vectors in AID systems}
\label{sec:attack}
Building on the discussion of security vulnerabilities in AID systems, it is crucial to examine the specific attack vectors that target these systems. Table \ref{table3-2} presents a reverse chronological overview of real-world attacks and emerging threats identified in the literature, detailing their causes, impacts on device functionality, and the adversarial capabilities required.

As illustrated in the taxonomy (Figure \ref{Taxonomy}), malicious attacks on AID systems can be broadly categorised into cyber-physical attacks and process-aware attacks, based on whether they exploit knowledge of the system’s dynamic control processes \cite{144}. Cyber-physical attacks, which are common across various IoMT devices, compromise system functionality by targeting software, hardware, or communication channels without requiring an in-depth understanding of the underlying control mechanisms. In contrast, process-aware attacks specifically target semi- or closed-loop control like AID systems. By making subtle manipulations within the operational bounds, these attacks can significantly impact system behaviours to threat patient safety \cite{2}. Since cyber-physical attacks align with broader IoMT security concerns, threats and potential countermeasures have been detailed in existing surveys \cite{82}\cite{93}\cite{94}. This section focuses on analysing attack vectors and technical details that are specific to AID systems. 

For systematic analysis, the attack vectors are grouped into three categories corresponding to the fundamental principles of cybersecurity: confidentiality, integrity, and availability. Table \ref{table3-1} provides this classification and links the attacks identified in Table \ref{table3-2} with feasible countermeasures summarised in Table \ref{table4-1}. The following discussion focuses on these attack vectors and outlines potential countermeasures, whereas Section \ref{sec:defence} reviews research contributions on implementing such defences in AID systems.

\input{Table_3-2}

\input{Table_3-1}

\input{Table_4-1}

\subsection{Threats to Confidentiality}
Ensuring the confidentiality of diabetes devices aims to guarantee that sensitive information is accessible only to authorised users or systems. Threats to this security aspect can originate from unauthorised individuals or even from patients themselves. 

\subsubsection{Eavesdropping} 
This refers to the unauthorised interception of wireless communication, which can expose personal patient information related to therapy and physical condition,  device information, and the device's personal identification number (PIN). Laboratory experiments have shown that data transmitted by a commercial insulin delivery system without cryptography can be intercepted using off-the-shelf Universal Software Radio Peripheral (USRP) and free software \cite{1}\cite{41}. By capturing its identification number, attackers could then gain full control of the pump. At a 2012 security conference, an engineer showed that a fatal dose of insulin could be delivered from 300 feet away  \cite{103}. These attacks exploit wireless communication vulnerabilities, such as wide transmission range and inadequate encryption. To mitigate these risks, countermeasures proposed in the literature mainly focus on (1) enhancing data protection through cryptography and secure communication protocols, and (2) limiting transmission range by employing proximity communication channels.

\subsubsection{Do-it-yourself (DIY) hacking}
This refers to unauthorised access or modification of commercial diabetes devices by patients themselves, driven by the desire for better data visualisation and more advanced control with shorter official censor and approval time \cite{15}.  For instance, Nightscout is an open-source project that helps patients hack their Dexcom CGMs for remote glucose level monitoring on mobile devices. Projects such as \#DIYPS enable patients to control insulin delivery. While these initiatives empower patients, they introduce risks due to vulnerabilities in unauthorised software and users' potential lack of expertise. Compared to FDA-regulated medical devices, these systems lack rigorous testing and responsible parties to support them in case of problems. To balance the need for patients to have greater control with secure AID systems against malicious exploitation, it is crucial to implement robust authentication strategies.

\subsection{Threats to Integrity}
The integrity of AID systems ensures that their data, algorithms, and operations remain accurate and consistent. Integrity can be compromised through two key aspects: data manipulation and algorithm manipulation. 

\subsubsection{Manipulation of Data}
Attackers can manipulate wirelessly transmitted data, such as CGM readings or controller commands,  to mislead the control algorithm or insulin pumps into issuing incorrect insulin delivery. Data manipulation can occur through the following attack vectors: 
\begin{itemize}
    \item Replay attacks: These attacks involve intercepting and replaying legitimate data packets to manipulate insulin delivery \cite{7}\cite{142}. This type of attack is possible when systems lack freshness checks or authentication, and attackers can observe and record communications. It does not require device access or message decryption. 
\end{itemize}

Once the attacker gains access to the device (e.g., by accessing the PIN) and decrypts the message, they can craft forged data packets to be accepted by the device  \cite{1}, enabling bias injection and false data injection attacks.

\begin{itemize}
    \item Bias Injection Attacks: In this attack, the attacker skews sensor readings by injecting a bias   \cite{42}.   By making subtle, consistent changes, the attacker can cause the system to under- or over-deliver insulin without needing extensive knowledge of the patient model or system operations. They can be difficult to detect due to their subtlety and correlation with the original data \cite{43}.  
\end{itemize}

\begin{itemize}
    \item False Data Injection Attacks (FDIA): FDIA refers to an adversary deliberately manipulating measurement or communication data to affect system state estimation or control decisions. They can be broadly categorised into random and targeted types, depending on how the data is forged. Random FDIA introduces arbitrary perturbations into input data. Experiments in \cite{44} showed that such attacks can shift the outputs of deep learning (DL) based glucose prediction models by 10-20 mg/dL under different scenarios. These attacks are generally easier to detect using sequential change detection techniques.
    
    In contrast, targeted FDIA carefully crafts input values to evade alarms and cause deliberate harm. This belongs to process-aware attacks that require knowledge of the control algorithm, which is often detailed in documentation. These attacks can be divided into model-specific and patient-specific variants. Model-specific FDIA exploit vulnerabilities in the target model; for example, Elnawawy et al. \cite{45} used URET to craft semantically and functionally plausible adversarial data points for interference AI models, while Kashyap et al. \cite{44} formulated this attack as an optimisation problem, showing that minimal, hard-to-detect perturbations against DL-based models can be computed within 60 milliseconds. Patient-specific FDIA endangers selected individuals, such as personalised insulin dose manipulation attacks craft attacks that harmful for target patients and within safety constraints \cite{10}, register false states (e.g., meal intake) to covertly mislead the control logic \cite{2}, or remote control insulin delivery \cite{1}\cite{100}\cite{41}.

\end{itemize}

Defence strategies for these data manipulation attacks typically include developing intrusion detection systems to detect abnormal patterns in sensor data and assessing controller activities. Another defensive measure involves using additional sensors or software monitors to cross-check data sources for inconsistencies. 

\subsubsection{Manipulation of Algorithms}
Algorithm manipulation involves tampering with the computational logic or parameter settings of AID systems, targeting either control algorithms or physiologic models used for patient state prediction  \cite{4}.  These include both traditional rule-based algorithms and machine learning(ML)-based models.

\begin{itemize}
    \item Computational attacks: These attacks may compromise instructions or internal data in algorithm execution \cite{4}. For example, Aliabadi et al. \cite{40} identified that the synchronisation between CGM and controller could be manipulated by altering the timing function  (get-data-timer()). This modification allows attackers to delay or drop critical messages.  Other threats involve tampering with essential device settings, such as disabling pump alarms, modifying the maximum dosage limits \cite{100} or pump I/O operations \cite{81}.
    
    \item ML model attacks: ML models are increasingly used in AID systems to enhance performance \cite{135}. For instance, the FDA-approved One Drop Blood Glucose Monitoring System employs a multi-layer perceptron (MLP) model for glucose prediction. Deep reinforcement learning (RL) approaches have been explored for use in AID systems for BG control \cite{141}. However, due to their black-box and data-dependent nature, understanding and monitoring the reasoning behind these models is challenging.  Attackers may influence the model’s behaviour during training or inference phases or reverse-engineer the model to extract confidential information \cite{45}. Moreover, ML-based algorithms can be severely interfered with by a very slight input data perturbation that is hard to detect—an attack variant described as ripple FDIA \cite{44}. In \cite{44},  minimal interference was calculated to misguide the ML model into predicting a patient’s glucose level as hyperglycaemic, leading the controller to administer an excessive dose of insulin. Experiments in \cite{140&145} show that a small perturbation around ±1mg/dL on BG reading can lead to severe patient harm over time in an RL controller model.
\end{itemize}

To defend against algorithm integrity threats, formal verification techniques are often employed to assess the operations of rule-based algorithms.  For ML-enabled systems, where the black-box nature poses challenges for formal verification, intrusion detection systems and modelling insulin injection patterns can be used to detect abnormalities and inconsistencies.

\subsection{Threats to Availability}
Availability in the context of AID systems refers to ensuring that the system and its components remain reliably functional for data transmission, processing, and regular operations. Threats to availability potentially lead to failures in glucose monitoring or delays in insulin delivery. Key threats include:

\begin{itemize}
    \item Denial-of-Service (DoS) and Distributed DoS (DDoS): In these attacks, an adversary floods the system with excessive traffic, overwhelming AID systems and preventing legitimate requests from being processed promptly \cite{81}. This could lead to delays or complete failure in delivering insulin \cite{73}.
    \item Ransomware Attacks: Ransomware can encrypt data and lock critical components of AID systems, demanding ransom payments to restore access \cite{48}. This can prevent patients or healthcare providers from viewing essential data, such as glucose levels or delivery history, rendering the system inoperable until resolved. 
    \item Jamming Attacks: By interfering with the wireless communication frequency, attackers can interrupt data flow, causing the system to fail to deliver insulin as required \cite{158}.
    \item Firmware/Software Corruption: Malicious firmware or software corruption can cause AID systems devices to malfunction or stop functioning entirely \cite{48}. For example, saturation-based sensor spoofing attacks make a sensor ignore legitimate inputs by using an additional infrared source to saturate the sensors \cite{146}. Additionally, attacks may exploit software vulnerabilities or delivering harmful updates can prevent systems from operating as intended, leading to a cessation of insulin administration. 
    \item Routing Protocol Information Attacks: Manipulating data transmission pathways in networked AID systems, attackers can disrupt or delay communication between key system components, compromising timely insulin delivery and glucose monitoring \cite{99}.

\end{itemize}

Mitigation strategies for these threats include network traffic monitoring, secure communication protocols, authentication mechanisms, formal verification, and redundant hardware or software components.

\section{Defence Strategies for AID systems}
\label{sec:defence}
Addressing the wide range of potential cyberattacks on AID systems requires comprehensive defence strategies that span multiple system layers.  Defence mechanisms in AID systems can be grouped into three primary areas: protected communication, intrusion detection systems, and control strategies assessment. Besides these internal defences, external strategies—such as the use of redundant hardware or software components—are employed to detect abnormal activities independent of AID devices. Table \ref{table4-1} provides an overview of the defence types and corresponding methods proposed in past research.

\subsection{Protected Communication}
Cyberattacks often exploit vulnerabilities in wireless communication channels. A variety of defence mechanisms have been proposed to enhance the security of information transmission within AID systems, including advanced authentication methods, encryption techniques, secure communication protocols, and specialized communication channels. 

\subsubsection{Authentication Methods}
Reliable authentication is crucial for ensuring secure communication within AID systems and for only allowing remote control with trusted devices.  Several authentication methods have been explored to meet the high-security and low computational demands of AID systems.

Biometric authentication methods like fingerprint recognition \cite{52}\cite{55} and voiceprint analysis \cite{56} are user-friendly and offer strong uniqueness. However, they require active user intervention, making them more appropriate for infrequent control operations (e.g., bolus injections) rather than continuous biosignal transmissions that occur periodically (e.g., every 5 minutes). Moreover, they raise concerns regarding the secure storage of sensitive biometric data, which could be spoofed by advanced attacks. To enhance system autonomy and reduce the risk of key leakage, hardware-based security methods such as Physically Unclonable Functions (PUFs) exploit inherent and unique physical variations in manufacturing processes, offering low computational overhead and an unclonable security mechanism without the need for key storage \cite{68}\cite{70}. Additionally, blockchain technology has been explored to protect patient data by decentralising data storage in central cloud servers \cite{68}. Rahmadika et al. \cite{75} further incorporated blockchain with federated learning for sharing local parameters for the misbehaviour detection model in AID systems without compromising patient privacy. 

\subsubsection{Encryption Techniques}
Following successful authentication, establishing encrypted communication between AID systems components is vital for maintaining data confidentiality and integrity. A key challenge is the energy and computational constraints. High-complexity encryption algorithms can introduce excessive computational overhead, reducing transmission frequency and accelerating battery depletion—which could hinder the system’s ability to respond promptly to critical scenarios. Another challenge is key distribution. While private key encryption enhances security, it may prevent medical professionals from accessing patient data in emergencies \cite{6}. 

To address these challenges, several studies have proposed encryption methods specifically tailored for AID systems. Table \ref{table4-2} summarises these approaches, detailing their methodologies, key sizes, security strengths, and limitations.

\input{Table_4-2}

Li et al. \cite{1} first demonstrated the vulnerability of wireless communication between glucose monitors and insulin delivery systems due to the lack of standard cryptography; they proposed using rolling code-based encryption—similar to that used in garage door systems—to prevent eavesdropping and replay attacks. However, the security of rolling code methods does not offer strong authentication and is highly dependent on encryption and decryption algorithms. Some methods, such as Keeloq, have been successfully compromised  \cite{147}. To enhance security, Marin et al. \cite{41} employed Advanced Encryption Standard (AES), a more robust symmetric encryption algorithm. Additionally, they reduced message size and transmission frequency to optimize computational efficiency. Their experiments also revealed that authentication for messages consumed more energy than encryption due to the need to append a message authentication code (MAC), which inevitably increased message length. A more lightweight solution proposed by ForkAE \cite{100} combined encryption and authentication processes to reduce computational overhead. 

Both AES and ForkAE rely on symmetric encryption, where a public key is used for both encryption and decryption. However, this approach carries the risk of key interception. To mitigate this problem, asymmetric encryption was introduced in \cite{52}, using a private key in every packet. While this approach enhances security, it requires larger key sizes, resulting in higher computational costs and vulnerable to factorisation attacks. To further protect patient privacy during computation, Weng et al. \cite{72} proposed using homomorphic encryption within a PID-based controller. This approach enables secure data processing without key sharing, as it allows computations to be performed on encrypted data. However, the CKKS homomorphic encryption method used in their study does not support division and comparison operations, potentially leading to imprecise computational results. Moreover, its high computational demands limit its application to more complex controller algorithms.

\subsubsection{Secure Communication Protocols}
A communication protocol defines the rules and standards for data exchange between devices, including authentication and encryption methods, to ensure secure and efficient transmission. Traditional high-security communication protocols often fall short in AID systems due to their substantial computational and energy demands. In contrast, low-power communication protocols like Bluetooth Low Energy (BLE), commonly used in IoT devices, often lack security robustness. 

When applied in AID systems, the absence of robust authentication mechanisms in BLE makes it vulnerable to man-in-the-middle and replay attacks. To address this concern in AID systems, Kwon et al. \cite{155}  proposed a protocol that introduces mutual authentication with 35.7\% less computational overhead compared to BLE. Kim et al. \cite{77} proposed a security protocol dedicated to AID systems by establishing tripartite authentication and a secure channel which is more resilient in emergencies. Except for authentication, the original protocol of a CGM (Dexcom G4) has been demonstrated to lack encryption \cite{158}, leaving it susceptible to eavesdropping and integrity attacks. To mitigate these threats, Reverberi and Oswald enhanced its protocol by incorporating AES-128 encryption and authentication, and introduced random keys and frequencies to protect against jamming attacks \cite{158}.  While this approach mitigates simple narrow-band jamming, it remains susceptible to high-power wide-band or more advanced jamming techniques. Strydis et al. \cite{48} introduced an energy-efficient protocol that runs on a separate core powered by RF-harvested energy until it performs external reader authentication.  This approach reduces the energy burden and enhances the system's resilience to battery DoS attacks.

\subsubsection{Proximity Communication Channels}
Besides protecting the data transmission process, limiting transmission range through proximity-based communication can reduce the risk of data interception and manipulation.  Li et al. \cite{1} proposed hiring body-coupled communication (BCC) that uses the human body as the wireless communication medium, which reduces wireless interception risks and power consumption. Limitations include that BCC can be affected by variations in body conductivity and movement. An alternative is visual light communication \cite{53}, which transmits data over short distances using visible light. This approach makes interception difficult and prevents remote attacks from hidden locations, but is sensitive to obstacles and ambient light, limiting its use to controlled environments like hospitals.

\subsection{Intrusion Detection Systems}
Monitoring and analysing system activity and states is vital to detect any intrusion and abnormal behaviours, ensuring the proper functioning of AID systems. Intrusion Detection Systems (IDS) have emerged as a key defence mechanism to counter threats. These systems are classified into three primary types: signature-based, specification-based, and anomaly-based approaches. Details of the intrusion detection system developed for AID systems—including proposed methods, targeted threats, experimental datasets, and performance metrics—are summarised in Appendix \ref{app:ids}.

\subsubsection{Signature-based IDS}
Signature-based IDS detects malicious activities by comparing real-time system behaviour against a database of known attack patterns. For example, Zhang et al. \cite{6} detect malicious data transactions in AID systems by defining abnormal physical signal characteristics and behaviours of the underlying content. The patterns of replay and DoS were defined in \cite{73}, and computational, data integrity, and communication attacks were defined in \cite{4}. For unintended failure detection, Howsmon et al.\cite{149} hired patient-specific metrics to monitor infusion site failures, and Baysal et al. \cite{160} defined sets of rules for detecting pressure-induced sensor attenuations. 

One key advantage of signature-based IDS is its lightweight computational overhead, making it well-suited for resource-constrained medical devices. However, it is ineffective against zero-day (unknown) attacks. Besides, updating signature databases in AID systems presents a challenge due to the stringent regulatory approval processes for medical devices. Despite these limitations, signature-based IDS can serve as a first-line monitor by rapidly filtering known threats and combining with other IDS capable of detecting unknown attacks to offer more robust defences.

\subsubsection{Specification-based IDS}
Specification-based IDS identifies malicious activities by comparing system behaviour to predefined rules or models that outline the expected operations of the device. Several studies have been conducted to improve formal specifications for AID systems, aiming to enhance precision and efficiency.

To mitigate the high resource demand of specification-based IDS for real-time performance,  Chen et al. \cite{159}  proposed a simplified multi-basal insulin infusion control model for verification, which addresses time-triggered jumps and merges flow pipes over the same time interval.  Alshalalfah et al. \cite{109} integrated statistical analysis with formal methods, employing statistical model checking to minimize exhaustive state exploration.  Subsequently, they proposed a dynamic specification mining approach using Bayesian networks to refine rules, reducing computational load while maintaining detection accuracy \cite{60}. Khan et al.\cite{4} introduced a declarative property-based approach, specifying only the initial and final states rather than all intermediate states in a runtime monitor. 

Rigid thresholds present another issue, as they may not adapt well to individual patient variability. Zhou et al. \cite{5} tackled this by personalising thresholds through adversarial training on fault-injected data, enabling more precise monitoring. Beyond internal system states, unsafe AID system operations can manifest as external abnormalities, such as repeated bolus requests or mismatches between insulin delivery and meal intake. Prematilake et al. \cite{61} developed a rule-checking mechanism that monitors both extrinsic and internal states. 

An ideal specification-based IDS is effective in defending against unknown attacks. However, constructing comprehensive rule sets to capture the complex and dynamic interactions within closed-loop insulin delivery systems, while limiting computational overhead, remains a challenge in practice. 

\subsubsection{Anomaly-based IDS}
Anomaly-based IDS identifies deviations from expected data patterns to detect abnormal activities. These approaches are generally categorised into model-based and model-free methods.

Model-based approaches consider a T1D patient as a dynamic system,  where the glucose concentration $g(t)$ is influenced by the insulin injection $i(t)$ and process disturbances $m(t)$ (e.g., meals and other patient activities). Intrusions are detected by comparing model predictions with observations. Based on the type of model, these approaches are further split into physiological model-based and data-driven model-based methods.

\textbf{Physiological model-based methods} utilise mathematical models to represent T1D patient profiles, often integrating predictors such as Kalman filters for state estimation. Early work by Vega-Hernandez et al. \cite{46} leveraged Hovorka’s nonlinear predictive model  \cite{104}, while Avila and Martínez \cite{163} combined the Lehmann and Deutsch model with a stochastic process to detect insulin pump issues. 

Several studies have focused on improving the prediction accuracy of these models. Tosun et al. \cite{74} added sequential change detection mechanisms to improve detection speed and proposed a KL-divergence bias-sensitive filter to improve the detection rate of constant bias injection attacks \cite{162}. To distinguish unregistered meals that caused false alerts, they introduced an online meal estimator using $\chi^2$ test with adaptive thresholds \cite{42}, which was later enhanced by including the statistical evaluation of prediction error \cite{79}. A comparison of three nonlinear predictive filters for online drift detection in BG data was presented in \cite{51}. 

The advantages of physiological model-based approaches include the use of expert knowledge, explainability, and a reduced requirement for extensive patient data. However, their accuracy is highly dependent on the fidelity of the underlying model and patient parameters, which at most can only partially approximate the complex and dynamic physiological processes of the human body.

\textbf{Data-driven model-based methods} rely on learning glucose-insulin dynamics from collected data. Early methods primarily employed statistical or linear models to represent these dynamics. For example, multivariable statistical monitoring methods were used in \cite{164} for detecting faults in glucose concentration values. Facchinetti et al. \cite{47} proposed individualized linear models to improve glucose predictions at night, which Del Favero et al. \cite{49} later expanded to cover day-long monitoring by incorporating meal effects. To distinguish between unannounced meal interference and pump malfunctions, they introduced two different sets of parameters in a personalized autoregressive moving average model \cite{168}. This work was further enhanced by incorporating a comprehensive analysis of various statistical properties associated with the residuals \cite{167}.

In recent years, ML and DL methods have been applied to model glucose-insulin dynamics. For instance, Olney et al. \cite{69} employed nonlinear autoregressive neural networks to detect and correct errors in CGM data, while Mahmud et al. \cite{80} individualized this model to enhance prediction performance. Rahmadika et al. \cite{75} developed a BiLSTM-based two-tier system for anomaly classification based on predictive deviations. Beyond modelling glucose-insulin dynamics, Maity et al. \cite{76} proposed to model the system’s operational characteristics and detect attacks by periodically relearning the model and comparing coefficients. 

Though high detection accuracy is achieved with these models, they often lack explainability and may not perform well on out-of-distribution data. To address these limitations, Venugopalan et al. \cite{81} suggested complementing ML-based models with personalised physiological models to enhance robustness. Zhou et al. \cite{67} incorporated unsafe action specifications into the training process using semantic loss regularisation. Building on this approach, they later introduced a data-driven hazard mitigation method that uses the desired sequence of recovery states to generate a corresponding sequence of control actions \cite{166}.

A key challenge faced by model-based IDS is that their performance depends on the accuracy of the patient model. However, the model highly varies between patients and fluctuates over time \cite{5}. Fixed models that do not account for these dynamic changes will struggle to maintain consistent performance over time. This underscores the need for personalised models that can continuously adapt to patients’ evolving physiological state to ensure reliable and effective anomaly detection. 

\textbf{Model-free anomaly detection methods} leverage data-driven classification techniques to identify abnormal patterns without requiring a patient-specific physiological model. These methods are divided into supervised and unsupervised approaches based on the need for labelled abnormal data. 

\begin{itemize}
    \item Supervised Learning: These methods train classifiers on labelled datasets to recognise faulty or attacked data. Rojas et al. \cite{58} mapped labelled data onto a feature map and defined boundaries for the “faulty” class, flagging patient glucose trends that enter this region. Judith et al. \cite{78} employed a multilayer perceptron model for classifying biometric data and network flow metrics. Despite their effectiveness, supervised methods face challenges due to the rarity and difficulty of obtaining labelled faulty or attacked data in the real world.

    \item Unsupervised Learning: To address the lack of labelled data, Meneghetti et al. \cite{161} 
    used 2 unsupervised anomaly detection methods: Local Outlier Factor and Connectivity-based Outlier Factor. Then they evaluated several unsupervised algorithms with feature selection techniques on simulated AID systems malfunction data \cite{54}. Later, they extended this work to real patient data, enhancing the differentiation between normal and abnormal by adjusting the data segmentation \cite{57}. Their follow-up study used a larger clinical dataset and adapted features for cases without meal announcements  \cite{9}.  Herrero et al. \cite{65} demonstrated that training a support vector machine to classify CGM data streams could accurately identify individual patient data. This method potentially aids in the detection of compromised data by reidentification.

\end{itemize}

While model-free methods bypass the need for profiling a physiological model, they still face challenges like managing false positives, optimizing feature selection, and handling data distribution shifts to maintain effective anomaly detection in AID systems.

\subsubsection{Hybrid methods}
Hybrid methods combine various IDS approaches to enhance detection accuracy and robustness. For instance, Astillo et al. \cite{71} merged model-based and model-free approaches, using deep learning models to forecast glucose levels and classify anomalies. They also incorporated outlier detection into specification rules to capture deviations in glucose data \cite{59}\cite{64}. Venugopalan et al. \cite{81} combined formal verification of critical system components with machine learning for insulin prediction, applying patient-specific safety thresholds. These approaches leverage the precision of specification-based or model-based methods and the adaptability of machine learning for more robust AID systems security. 

\subsection{Control Strategies Assessment}
To prevent abnormal insulin delivery resulting from compromised CGM data, controller algorithms, or insulin administration, various defence strategies have been developed to evaluate and verify insulin dose control mechanisms. These strategies can be categorized into rule-based, data-driven, and hybrid approaches. 

\subsubsection{Rule-based methods}
Rule-based methods compare insulin doses against predefined thresholds or patterns for anomaly detection. Khan et al. \cite{4} used static threshold ranges to detect overdoses or underdoses, which, while effective for extreme deviations, fail to capture sophisticated attacks that account for patient variability. To improve this, Zhou et al. \cite{5} introduced personalised thresholds and specific abnormal behaviours to align control actions with patient needs. However, this method detects anomalies after the patient’s physiological response to the attack becomes apparent, leaving less time to prevent harm, especially in the case of underdosing, where adverse effects manifest more subtly.

\subsubsection{Data-driven methods}
Data-driven methods detect anomalies by predicting insulin doses using historical data, aiming to enhance timeliness and accuracy. Hei et al. \cite{3} employed supervised learning to model normal infusion dosages and rates against time. However, their approach assumes rigid patient routines, limiting applicability, especially in pediatric cases. Levy-Loboda et al. \cite{10} exploited more data sources such as the logs of the insulin pump, CGM, and relevant medical information. They simulated insulin dose manipulation attacks and mined temporal patterns to identify anomalies using Logistic Regression and Random Forest. However, their study focused only on nighttime basal injections, neglecting other activities like meals and bolus doses. 

A significant gap remains in the development of comprehensive models that integrate patient activities, meal intake, and physiological dynamics to enable accurate and robust profiling of insulin delivery patterns.

\subsection{Redundant Hardware or Software Components}
External defences employ redundant hardware or software components to detect abnormalities without adding computational strain to AID systems, allowing for independent updates without recertification for the medical device. For hardware auxiliary components, Strydis et al. \cite{48} designed an energy-efficient communication protocol that runs security functions separately from core medical operations, minimising interference with the primary system. MedMon \cite{6} utilised a USRP device to intercept and analyse the signal and content features of remote controller commands. user authentication was required if abnormalities were detected, thereby enhancing security. This approach demands fast computation speed and may require frequent user interaction. Panda et al. \cite{62} integrated a wearable ECG sensor to detect anomalies by correlating ECG intervals with blood glucose levels, improving system robustness but facing challenges in accurately correlating data across varied patient conditions. For software auxiliary components,  an additional safety layer for control algorithms has been proposed to prevent the controller from violating constraints with medical criteria \cite{165}. A virtual auxiliary system with detection filters was introduced to identify replay cyberattacks \cite{7}.

While these external defences provide added protection without affecting device performance, they may be vulnerable to initial compromise and can be combined with internal security measures for comprehensive AID systems protection.

% --------------------------------------------------------------------------

\section{Experiment Settings and Available Resources}
\label{sec:resources}
Evaluating security systems for AID systems requires datasets that capture both the effects of cyberattacks on device functionality and the resulting physiological changes in patients. However, collecting such data through experiments is impractical due to safety concerns. Besides, clinical data involving suspicious behaviour linked to cyberattacks is rare and difficult to label. These limitations often necessitate the use of simulation tools for testing and validating defence mechanisms. This section reviews evaluation methods and key experiment resources, including hardware, simulators, and datasets used in AID systems security research. Table in Appendix \ref{app:data} summarises the key experimental resources, detailing their descriptions, available sources, and relevant applications.

\subsection{Hardware Simulation}
Hardware simulations offer detailed insights into physical aspects such as communication, energy consumption, and heat dissipation, which are crucial for assessing the real-world application of security measures. These simulations focus on evaluating system-level vulnerabilities and performance for defence at the communication and external levels. 

Commercialised connected diabetes devices are key resources for hardware simulation. These devices serve a dual purpose: demonstrating cybersecurity weaknesses and validating defence mechanisms. For example, Li et al. \cite{1} and Marin et al. \cite{41} showed that radio communications in commercial AID systems can be intercepted and compromised using a USRP. Conversely, Zhang et al. \cite{6} used a USRP as the safety monitor for wireless communication, and Lazaro et al. \cite{52} tested a communication protocol by combining a commercial CGM, insulin pump, and control algorithm. 

Some studies focus on improving the hardware or software of AID systems to enhance security and validate the efficiency of their methods. These studies often use more general and flexible hardware boards to replicate the system. For instance, Strydis et al. \cite{48} used Biostator II, an ultra-low-power ASIC controller, to evaluate the energy efficiency in the proposed security communication protocol. Zhao et al. \cite{53} simulated on an Arduino Uno-R3 microcontroller to assess a visible light access control channel. The Arduino platform was also employed for hardware-in-the-loop simulation to test controller algorithms \cite{170} and simulate the responses of cyber-attacks \cite{169} in AID systems. Prematilake et al. \cite{61} implemented a safety-enhanced insulin pump system on the EFM32WG to monitor both external and internal states. Kim et al. \cite{77} tested an AID systems-specific secure communication protocol using an Arduino Nano 33 for the insulin pump, an nRF52840 kit for the CGM, and a Galaxy S20 for the controller. Ahmed et al.\cite{127}  developed a hardware emulation platform using an ESP8266 kit to emulate faults, attacks, and test countermeasures for evaluating closed-loop control systems. 

In summary, by leveraging commercial medical devices and custom hardware setups, researchers can identify cybersecurity weaknesses and assess the real-world impact of security measures on AID systems' performance. 

\subsection{Software Simulation}
Software simulation, also known as in silico trials, plays a crucial role in evaluating closed-loop insulin delivery systems by mimicking patient responses to various controller strategies and potential cyberattacks. This approach enables researchers to simulate how attacks targeting CGM readings or insulin delivery commands affect the human body and assess the effectiveness of defence mechanisms at the intrusion detection and control strategies evaluation level. Closed-loop AID testbeds typically integrate T1D patient simulators that combine pharmacokinetic models with meal and activity simulations, and controller algorithms.

Pharmacokinetic models are mathematical representations that describe the absorption, distribution, metabolism, and elimination of drugs (e.g., insulin or glucagon) in the body. They are primarily used to predict drug concentration-time profiles and optimise dosing regimens. For example, Bagade et al. \cite{107} utilise a hybrid physiologically-based pharmacokinetic model \cite{106} to assess the computational overhead of controller algorithms running as Android services and Vega-Hernandez et al. \cite{46} leveraged Hovorka’s nonlinear predictive model (i.e., Cambridge model) \cite{104} for state estimation. Beyond simulation, these models have also been incorporated into physiologic model-based intrusion detection systems to support prediction and anomaly detection \cite{51}\cite{42}. 

T1D patient simulators are integrated platforms that model the entire glucose-insulin regulatory system, often incorporating physiological variability and patient behaviours (e.g., meal timing, insulin dosing decisions). They are used for in silico clinical trials, testing new therapies, and simulating real-life scenarios. One widely used model is the Padova/UVA T1D simulator, an FDA-approved nonlinear simulator for pre-clinical testing \cite{47}\cite{49}\cite{59}\cite{71}\cite{81}. This simulator captures interpatient variability by simulating 100 virtual subjects covering children, adolescents, and adults with varying physiological profiles. Since its initial release in 2008 \cite{112}, it has evolved to model patient physiology changes over time in 2013 \cite{34} and included intraday insulin sensitivity variations, time-varying therapy parameters, and a "dawn phenomenon" model in 2018 \cite{110}. Another simulator named Glucosym supports online simulations with 10 real patient profiles, used for attack simulations and defence evaluations in \cite{73}.  

Controller algorithms automatically calculate and adjust insulin delivery to maintain the patient’s glucose level within a target range. These algorithms can be integrated with patient simulators to model and evaluate AID systems' functionality. Common control strategies include Proportional-Integral-Derivative (PID) controllers, Model Predictive Controllers (MPC), fuzzy logic controllers, and other nonlinear control approaches. Notably, open-source algorithms such as OpenAPS \cite{117} enable T1D patients to connect commercial CGMs and insulin pumps, creating a hybrid closed-loop system for more automated self-management. A comprehensive review of patient models and control strategies for AID systems is provided in \cite{83}. 

To further streamline simulation, some studies integrate T1D patient simulators with controller algorithms to create closed-loop AID systems testbeds. For example, LoopInsightT1 \cite{LoopInsightT1} provides an in-browser simulation platform for modelling closed-loop interactions between individuals with T1D and AID systems. It integrates multiple physiological models, such as the UVA/Padova \cite{34} and the Cambridge model \cite{104}, with several control algorithms, and offers interactive visualisations to analyse diverse scenarios (e.g., the impact of meal announcements). Zhou et al. \cite{33} developed 2 closed-loop testbeds incorporating the UVA/Padova simulator with Basal-Bolus controller and Glucosym simulator with OpenAPS controller. They also provide a fault injection engine and a labelled dataset of simulated faulty records. Another available source is a MATLAB Simulink model testbed, which simulates and integrates a continuous glucose monitor, an insulin pump, and a PID controller \cite{122}. These integrated testbeds facilitate a more convenient and thorough evaluation of AID security.

\subsection{Real-world datasets}
Real-world datasets are important resources for evaluating the effectiveness of cyberattack defence systems. Although they often lack explicit attack data, they are valuable for assessing patient models and predictions, as well as the performance of proposed security mechanisms in real world.

Several datasets provide rich data for continuous glucose monitoring, both from free-living settings and clinical trials. Datasets like JCHR-DCLP3\cite{38}, REPLACE-BG \cite{118}, D1NAMO\cite{115}, and OhioT1DM Dataset \cite{126} provide CGM data that could be utilised for experiments involving glucose level prediction \cite{69}\cite{114} and re-identification \cite{65}. JCHR-DCLP3 dataset \cite{38} was employed by Zhou et al.  \cite{33} to test the performance of their proposed AID systems simulation testbeds, while Elnawawy et al. \cite{45} used the OhioT1DM Dataset \cite{126} to assess the impact of adversarial inputs on machine learning model predictions.

In addition to glucose monitoring data, some datasets provide supplementary biometric and context information. For example, the DICARDIA dataset \cite{116} contains ECG data, which Panda et al. \cite{62} utilised to develop safety monitors by analysing the correlation between ECG intervals and blood glucose levels. The WUSTL-EHMS-2020 dataset \cite{124}, though not diabetes-specific, offers network flow metrics and biometric data, proving useful for evaluating defence systems that monitor both biometrics and network anomalies, such as detecting man-in-the-middle attacks by Judith et al. \cite{78}. The OpenHumans dataset \cite{39}, collected and shared by a diabetes community, includes patient profiles, glucose levels, infusion doses, and ample activity records. Additionally, AID systems-faulty data, generated by the testbed described by Zhou et al.\cite{33}, comprises synthetic data from closed-loop simulations, including both safe and hazardous data traces from 20 patient profiles, providing a useful source for evaluating fault detection and risk management.

% -------------------------------------------------------------------------------
\section{Key Findings}
\label{sec:key_findings}
\subsection{RQ1: What potential attack vectors and risks do AID systems face?}

As summarised in Table \ref{table3-2}, the attack vectors targeting AID systems primarily threaten three core security dimensions: confidentiality, by intercepting sensitive patient data; integrity, by altering system data, commands, or logic; and availability, by disrupting the normal functions. 

Confidentiality attacks often utilise vulnerabilities in the wireless communication links, such as weak or absent encryption protocols, between system components, enabling adversaries to eavesdrop using off-the-shelf radio interception and generation devices. In some cases, such breaches may be initiated by patients seeking greater visibility or control over their treatment data. Availability attacks impair the system’s ability to function properly. DoS attacks, for instance, can interrupt the transmission of CGM readings and controller commands or deplete system resources by continuously sending malicious requests. These threats typically require moderate technical knowledge—communication protocols and device function—without necessitating a deep understanding of the system’s control process.

As physiological closed-loop control systems, AID devices are especially susceptible to integrity attacks, which can result in incorrect insulin dosing within a short time frame. These attacks may target any component in the loop, including CGM readings, food registration, control algorithms, controller commands, or the insulin pump driver. One of the relatively direct forms is the replay attack, where previously intercepted data is resent to the system. More advanced attacks involve injecting crafted false values— such as randomly generated or biased—to influence insulin delivery. These attacks require reverse-engineering the communication protocol to produce valid data packages  \cite{1}\cite{41}. A particularly sophisticated class of threats, known as process-aware attacks, involves leveraging the knowledge of the system's control process and states to manipulate the system into a harmful but acceptable state. Such as personalised insulin dose manipulation attacks that only threaten target patients in a specific condition, and ripple false data injection attacks that use minimal input perturbations to stealthily change the prediction of deep learning models in the controller \cite{140&145}.

\subsection{RQ2: How do existing defence mechanisms address these risks?}

Existing defence mechanisms proposed for AID systems operate across multiple layers, as summarised in Table \ref{table4-1}, targeting different stages of potential attack.

The first layer focuses on preventing wireless communication channels from interception and manipulation. Techniques such as authentication, cryptography, and secure communication protocols are employed to protect wireless transmissions between system components. Alternative communication approaches, such as body-coupled communication or visible-light communication, have also been proposed to minimise exposure to in-the-air communication threats. In addition, hardware-based measures, including wireless traffic monitors or dedicated computation cores, can be integrated to protect devices against availability attacks and ensure resilient communication. 

The second layer involves deploying intrusion detection systems to identify breaches that may have already occurred. These systems are designed to continuously monitor system activity for signs of compromise by analysing signatures of attacks, deviations from expected system operations, physiological states, or control behaviour patterns. This layer plays a crucial role in identifying emergent risks and enabling early-stage mitigation before any harmful operations occur.

The third layer aims to assess control strategies. In cases where attackers have successfully launched attacks to impact insulin delivery, this layer serves as a final checkpoint to identify hazards. It verifies whether the insulin delivery activity aligns with predefined safety constraints or conforms to the patient’s historical dosage patterns. By validating outputs at the control level, this layer mitigates the risk of incorrect or unsafe insulin administration.

An additional layer of protection involves redundant hardware or software components that operate independently of the main AID system. These auxiliary protections offer two key benefits. First, they avoid placing additional strain on the AID system’s internal resources, such as battery life, memory, or processing capacity. Second, because they function externally, they can be updated or patched more rapidly in response to evolving threats without requiring full medical device recertification. Examples include external monitoring modules for inspecting communication traffic and system data, or supplementary biosensors that provide physiological cross-verification.

\subsection{RQ3: How practical and robust are these solutions in protecting AID systems from cybersecurity threats?}

As outlined in Table \ref{table3-1}, most known threats map to corresponding defence strategies developed for AID systems. Yet, their robustness and real-world applicability vary with implementation constraints. The following discussion considers risks from primary categories to more nuanced threats, together with the defence strategies and requirements for practical deployment.

The most critical threats arise from weak wireless communication protocols that lack robust encryption and authentication, leaving systems vulnerable to eavesdropping and manipulation of transmitted data and commands. Early AID products' protocol lacks encryption and authentication, and multiple studies have demonstrated that they could be easily compromised. More recent systems now incorporate cryptographic and authentication mechanisms to protect wireless communication; however, their robustness depends on the strength of algorithms and key management practices. An implementation challenge is that highly secure protocols demand significant computational resources, which may impair system performance and reduce battery life. To address this, research has proposed adaptations such as reducing message size, lowering communication frequency, or encrypting critical modules (e.g., the controller) to increase security without adding much computation overhead. Authentication methods such as biometrics and hardware-based PUFs can be incorporated to further protect critical controller commands and biosignal transmissions from trusted sources. To defend against availability threats such as DoS attacks, additional hardware components—for example, dedicated traffic monitors or separate computation cores—can be integrated. These measures can be incrementally incorporated to strengthen existing wireless protections. While for highly security-sensitive or fixed environments (e.g., in-hospital use), proximity-based channels such as body-coupled communication or visible-light communication offer promising alternatives to avoid over-the-air exposure.

Even the strongest secure communication protocol cannot guarantee complete protection. If adversaries manage to compromise the closed-loop control process, monitoring both physiological and system states becomes essential for detecting integrity attacks and preventing hazards. Security monitoring in current commercial products is relatively basic, relying primarily on threshold-based mechanisms to flag extreme values for BG, insulin delivery dosage, and injection rate. While effective at identifying attacks that cause extreme deviations (e.g., delivering a maximum insulin dose) under the assumption that monitored values are uncompromised, these mechanisms fall short in addressing more subtle and covert threats. For instance, replay attacks or targeted false data injection attacks, which alter glucose readings or commands within limits, may evade threshold-based detection. Sequential change detection techniques can help identify discontinuities in BG readings. For attacks that craft gradual changes, signature-based IDS can rapidly filter known attack patterns, while specification-based IDS can validate controller behaviour against system specifications to ensure compliance with expected operation. Collectively, these defence strategies are broadly applicable for effectively detecting many known threats and activities that deviate markedly from normal device operation.

More nuanced threats, such as process-aware and personalised attacks, can manipulate data in ways that cause targeted harm without violating system specifications or matching known attack patterns. Detecting such threats requires more sensitive monitoring approaches that model the patient’s physiological state and insulin delivery patterns. By identifying deviations from an individual’s normal glucose or insulin profiles, these potential attacks could be detected. Another method is the use of multiple biometric sensors worn by the patient, where correlations across signals (e.g., ECG and BG measurements) are modelled to reveal anomalies. However, accurate modelling is difficult due to the inter-patient variability and changing physical conditions. Insulin delivery patterns also shift with context, lifestyle, and activity, meaning that models based solely on historical data are prone to false alarms once deployed. While advances in hardware, battery efficiency, and tiny ML will enable more capable models in the future, achieving a high-performance security monitor requires online, continuous learning without catastrophic forgetting, which remains an open challenge. Frequent offline retraining and updating may mitigate this issue, but are subject to medical device censor and unstable performance. 

Importantly, although launching such attacks requires a very high level of expertise for adversaries (i.e., break the communication protocol and possess a deep understanding of the control process and target patients’ unique characteristics), these advanced monitoring functions offer dual benefits: not only enhancing security by detecting data manipulations, but also capable of identifying device degrading performance or changes in patient physiological states. This dual capability holds significant potential for strengthening medical device reliability and enabling personalised healthcare (e.g., early detection of disease), shedding light on the development of future intelligent healthcare systems.

% -------------------------------------------------------------------------------
\section{Open Research Challenges and Future Directions}
\label{sec:challenges}
While the defence strategies discussed above potentially offer multiple layers of protection and monitoring, their real-world effectiveness depends on several critical factors that must be addressed when developing a comprehensive security framework for AID systems.

\subsection{Overall Security}
Although individual defence mechanisms target certain threats at different layers, a robust AID system needs to mitigate a broad range of attacks while ensuring timely responses to emergencies. This necessitates an integrated security architecture that combines diverse protective measures into a cohesive strategy. Key research directions include balancing security strength with the computational overhead of combined mechanisms and designing methods that complement one another to reduce computation redundancy while enhancing overall resilience.

\subsection{Resource Constraints}
As wearable medical devices, AID systems are constrained by limited computational power, memory, and battery life, which makes the deployment of resource-intensive security mechanisms particularly challenging. While specification-based and data-driven defences offer advanced security monitoring, implementing them without impacting device functionality remains an open problem. Future research should explore model optimisation, computation offloading, and hardware innovations to balance strong security with the operational constraints of AID systems.

\subsection{Variability Across Patients}
For monitoring of system and physiological states, a major challenge is distinguishing abnormal patterns from a patient’s normal behaviour, which requires accounting for both intra- and inter-patient variability. Each patient’s physiological responses, lifestyle, and treatment regimen are unique, making it difficult to develop universal monitoring models that adapt to these personalised characteristics. Future security monitors for AID systems need to dynamically accommodate patient heterogeneity while maintaining detection performance across the device lifecycle.

\subsection{Comprehensive Modelling of Insulin Infusion Patterns}
Accurately predicting insulin needs remains a challenge, particularly in response to dynamic contexts such as meal intake, physical activity, and stress. Current insulin infusion models struggle to account for the full range of variables affecting glucose levels. There is a need for comprehensive models that integrate multisource data to predict and adapt to patient-specific insulin requirements. Improving the accuracy of these models not only facilitates the development of fully closed-loop insulin control systems but also plays a critical role in defending against cyberattacks that manipulate injected doses within insulin delivery systems.

\subsection{Trustworthy Defense Mechanisms}
While securing AID systems is essential, it is equally important to ensure that defence mechanisms are reliable and preserve patient privacy. Many recent data-driven cybersecurity solutions operate as black boxes, providing little insight into decision-making, which complicates interpretation and intervention by users or healthcare professionals. Additionally, sharing data between devices to improve model accuracy introduces privacy concerns. Achieving effective knowledge sharing without compromising sensitive patient information remains a critical challenge, for which approaches such as federated learning may offer promising solutions.

\subsection{Standardisation of AID systems Cybersecurity Frameworks}
Current medical device cybersecurity standards and guidance leave the responsibility for comprehensive threat modelling and protective measures implementation with manufacturers, creating a gap between broad regulatory compliance and the satisfaction of the specific security needs of medical devices. AID systems, as physiological closed-loop medical wearables, face distinctive security and safety risks that directly affect drug delivery in real-time. Addressing this challenge requires complementary security frameworks that specify existing and emerging threats faced by AID systems and recommend
appropriate countermeasures and emergency responses to ensure system performance. Independent conformity assessments are also needed to build confidence among users and stakeholders. Ultimately, developing device-specific security requirements and assessment mechanisms will be essential for establishing more comprehensive and concrete regulations for medical device security.

\section{Conclusion}
This review provides a comprehensive analysis of the security landscape of AID systems, encompassing technical, regulatory, and commercial aspects. We examined existing attack vectors, defence mechanisms, and security assessment approaches in the literature, alongside available research resources. The priority of these threats and the practicality of corresponding countermeasures were evaluated, highlighting several critical challenges for future research. While substantial progress has been made, continued innovation and practical implementation are needed to develop thorough, resource-efficient, and patient-centric defence mechanisms. Establishing effective security measures to prevent, detect, and respond to threats in real time requires coordinated efforts among researchers, manufacturers, regulators, and healthcare providers. Successfully addressing these challenges is crucial for enhancing system reliability, building user trust, and fully realising the transformative potential of AID systems in improving diabetes care.

\appendices
\section{Table of regulation standards and guidelines of cybersecurity in medical devices}
\label{app:regulation}
\input{Table_2-1}

\section{Table of Cybersecurity in representative AID system products}
\label{app:products}
\input{Table_2-3}

\section{Table of intrusion detection systems developed for AID systems}
\label{app:ids}
\input{Table_4-3}

\section{Table of available resources and utilisation in AID cybersecurity research.}
\label{app:data}
\input{Table_5-1}

\section*{Acknowledgment}
This work was supported in part by the Ministry of Education, Singapore, under its Academic Research Fund Tier 2, under Grant MOE-T2EP20121-0008.

\section*{Declaration of generative AI and AI-assisted technologies in the writing process}
During the preparation of this work the authors used ChatGPT in order to improve the readability and language of the manuscript. After using this tool/service, the authors reviewed and edited the content as needed and take full responsibility for the content of the published article.

\bibliographystyle{IEEEtran}
\bibliography{main}

\end{document}

%% file: Table_1-1.tex
\begin{table*}[!h]
\label{table1}
    \centering
    \caption{Comparison of review papers addressing AID system security over the past decade. \\ {(Symbols:$\checkmark$ = covered, $\halfcheckmark$ = mentioned but lacks literature review and/or comprehensive coverage, $\times$ = not covered)}}
    \begin{adjustbox}{width=0.9\textwidth}
    \small
    \begin{tabular}{|>{\raggedright\arraybackslash}p{0.2\textwidth}|>{\raggedright\arraybackslash}p{0.3\textwidth}|>{\raggedright\arraybackslash}p{0.15\textwidth}|>{\raggedright\arraybackslash}p{0.15\textwidth}|>{\raggedright\arraybackslash}p{0.15\textwidth}|>{\raggedright\arraybackslash}p{0.15\textwidth}|}
    \hline
         Review Paper&  Focus &   Discussion on Unintended Failures&Discussion on Malicious Attacks&  Analysis of Attack Vectors&  Analysis of Protective Mechanisms \\
    \hline
         Klonoff, 2015 \cite{15}& Cybersecurity for Connected Diabetes Devices& $\times$& $\checkmark$& $\times$&$\halfcheckmark$ (No literature review)\\
    \hline
         O'Keeffe et al,. 2015 \cite{92}& Cybersecurity in AID systems& $\times$& $\checkmark$& $\times$&$\halfcheckmark$ (No literature review)\\
    \hline
         Blauw et al., 2016 \cite{91}& Safety and design requirements of the AID systems& $\checkmark$& $\times$& $\times$&$\halfcheckmark$ (Only for unintended failures)\\
    \hline
         Ramkissoon et al., 2017 \cite{89}& Specific hazards applicable to the AID systems & $\checkmark$& $\times$& $\times$& $\halfcheckmark$ (Only for unintended failures)\\
    \hline
         Thompson et al., 2018 \cite{29}& Safety issues in outpatient settings& $\checkmark$& $\checkmark$ & $\times$&$\times$\\
    \hline
         Quintal er al., 2019 \cite{88}& Ethical issues associated with AID systems & $\checkmark$& $\checkmark$& $\times$&$\times$\\
    \hline
         Singh and Rahman, 2020 \cite{87}& The security of Open AID systems system& $\times$& $\checkmark$& $\halfcheckmark$ (No literature review)&$\halfcheckmark$ (No literature review)\\
    \hline
         Sherr et al., 2022 \cite{150}& Current landscape and safety challenges of AID systems& $\checkmark$& $\checkmark$& $\times$&$\times$ \\
    \hline
         Wirth, 2022 \cite{154}& The states and tendency of cybersecurity in the diabetes ecosystem& $\times$& $\checkmark$& $\times$&$\times$\\ 
    \hline
         Nazzal et al., 2024 \cite{152} &  Software safety and security verification and validation on AID systems&   $\times$&$\times$ &  $\times$ &  $\halfcheckmark$ (Only formal verification methods) \\ 
    \hline
         Kapadia, 2024 \cite{153}&  Limitations and malfunctions&   $\checkmark$&$\times$ &  $\times$ &  $\times$ \\
    \hline
         This paper& The cybersecurity status, attack vectors, and 
         proposed defence strategies of AID systems& $\times$& $\checkmark$& $\checkmark$&$\checkmark$\\
 \hline
    
    \end{tabular}
    \end{adjustbox}

\label{tab:table1-1}
\end{table*}

%% file: Table_1-2.tex
\begin{table*}
    \centering
    \caption{Keywords used in search engines.}
    \begin{adjustbox}{width=1.0\textwidth}
    \begin{tabular}{|p{0.2\textwidth}|p{0.9\textwidth}|} \hline 
         Search Engine&  Keywords\\ \hline 
 Google Scholar& (intitle:"artificial pancreas" OR intitle:"closed-loop insulin pump" OR intitle:"automated insulin delivery" OR intitle:"hybrid closed loop systems")  AND (attack OR cyberattack OR hijack OR cybersecurity OR security OR "intrusion detection" OR secure)\\ \hline 
         Scopus&  TITLE-ABS-KEY ( ("artificial pancreas" OR "closed-loop insulin pump" OR "automated insulin delivery" OR "hybrid closed loop systems") AND (attack OR cyberattack OR hijack OR cybersecurity OR security OR "intrusion detection" OR secure))\\ \hline 
         IEEE Xplore&  ("artificial pancreas" OR "closed-loop insulin pump" OR "automated insulin delivery" OR "hybrid closed-loop systems") AND (attack OR cyberattack OR hijack OR cybersecurity OR security OR "intrusion detection" OR secure)\\ \hline 
         Web of Science&  ("artificial pancreas" OR "closed-loop insulin pump" OR "automated insulin delivery" OR "hybrid closed-loop systems") AND (attack OR cyberattack OR hijack OR cybersecurity OR security OR "intrusion detection" OR secure)\\ \hline 
         PubMed&  ("artificial pancreas" OR "closed-loop insulin pump" OR "automated insulin delivery" OR "hybrid closed-loop systems") AND (attack OR cyberattack OR hijack OR cybersecurity OR security OR "intrusion detection" or secure)\\ \hline
    \end{tabular}
    \end{adjustbox}
    \label{tab:table1-2}
\end{table*}

%% file: Table_3-2.tex
% \usepackage{multirow}
% \begin{landscape}

\begin{table*}[!htbp]
\centering
\caption{Documented malicious attacks on AID systems.}
\scriptsize
% \begin{adjustbox}{width=1\textwidth}
\begin{tabular}{|p{0.10\textwidth}|p{0.10\textwidth}|p{0.10\textwidth}|p{0.20\textwidth}|p{0.20\textwidth}|p{0.20\textwidth}|}
\hline
Attack Name &
  Reference &
  Source Type &
  Cause of Failure &
  Impact &
  Adversarial Capabilities \\ \hline
RL-based control model attacks &
  Chang et al. \cite{140&145} (10/2024) &
  Literature (Experimentally Validated) &
  The control system is vulnerable to anomalous inputs. &
  A maximum CGM reading of 400mg/dL caused high rates (over 40\%) of hypoglycaemic failure within a few hours; a very slight perturbation (±1mg/dL) can lead to disproportionate patient harm over time &
  High - Read access to the weights of the trained model or estimated weights \\ \hline
AI/ML model attacks &
  Elnawawy et al. \cite{45} (01/2024) &
  Literature (Experimentally Validated) &
  Introducing adversarial CGM data points during inference through a known vulnerability in the Bluetooth communication channel &
  Wrong diagnosis or treatment &
  Medium - Can exploit the Bluetooth communication stack to intercept and manipulate the CGM measurements \\ \hline
Unauthorised remote control &
  Medtronic \cite{143} (01/2023) &
  Real-world (Recall) &
  The settings and insulin delivery of the pump could be potentially controlled through a wireless connection nearby &
  Overdosing or Underdosing &
  Medium - Intercept and manipulate the communication between devices \\ \hline
Process-aware attacks &
  Stergiopoulos et al. \cite{2} (11/2023) &
  Literature (Experimentally Validated) &
  Utilising knowledge of the dynamic process to alter sensory data or the logic inside the controller in an intended but subtle way &
  Overdosing or Underdosing &
  High - Prior knowledge of the components’ implementation mechanism \\ \hline
Replay attacks &
  Walenda et al. \cite{73} (11/2023) &
  Literature (Experimentally Validated) &
  Sending repeating insulin dosages to the insulin pump through a vulnerability in the wireless protocol &
  Replay attacks have the greatest effect when the injected insulin doses are high &
  Low – Intercept the wireless connection and report past packets \\ \hline
DoS attacks &
  Walenda et al. \cite{73} (11/2023) &
  Literature (Experimentally Validated) &
  Overload the system to stop insulin delivery for extended periods &
  DoS attacks are most effective at low, increasing doses, causing longer out-of-range times and greater spikes and dips than replay attacks. &
  Low - Jam the communication channel \\ \hline
Personalised insulin dose manipulation attacks &
  Levy-Loboda et al. \cite{10} (08/2022) &
  Literature (Theoretical) &
  Craft attacks that only threat target patients and are capable of evading extreme overdose/underdose detection &
  Overdosing or Underdosing &
  High – Prior knowledge and manipulation of the insulin dosage \\ \hline
Bias injection attacks &
  Emre Tosunal \cite{42} (06/2022) &
  Literature (Theoretical) &
  A constant bias is injected into the compromised sensor readings &
  Overdosing or Underdosing &
  Medium - Intercept and   manipulate the communication between devices \\ \hline
Ripple false data injection attacks &
  Kashyap et al. \cite{44} (05/2021) &
  Literature (Experimentally Validated) &
  Using minimal input   perturbations to stealthily change the DNN-predicted insulin dosage &
  More allowed perturbations make attacks easier to find for targeted attacks, with discovery time as short as 60ms. Random attacks can easily induce 10–20 mg/dL output changes across scenarios. &
  High - Read access to the weights of the trained model or estimated weights \\ \hline
Command manipulation attacks &
  Khan et al. \cite{4} (12/2020) &
  Literature (Theoretical) &
  Change the instructions or   data that implement the algorithmic steps &
  Overdosing or Underdosing &
  High - Access to the relevant resources; modify a command (i.e., instructions that implement the   algorithmic steps) \\ \hline
Unauthorised remote control &
  Cooke et al.   \cite{100} (04/2020) &
  Literature (Experimentally Validated) &
  An adversary can remotely control insulin delivery and access unencrypted medical data &
  Overdosing or Underdosing &
  Medium - Intercept and   manipulate the communication between devices \\ \hline
Replay attacks &
  Medtronic \cite{142} (08/2018) &
  Real-world (Notification) &
  Involuntary bolus insulin could be delivered to   the pump user by copying the wireless radio frequency (RF) signals   from remote controllers in proximity &
  Overdosing &
  Low – Intercept the wireless connection and report past packets \\ \hline
 CGM attacks & Reverberi and Oswald \cite{158} (08/2017)& Literature (Experimentally Validated)& The Dexcom G4 protocol does not employ cryptographic algorithms & Enables the tracking of a user, the jamming of the communication channels and the forging of incorrect sensor readings&Low to medium - Software-defined radio and low-cost RF chipsets, reverse engineering\\\hline 
Saturation-based sensor spoofing attacks &
  Park et al. \cite{146}   (08/2016) &
  Literature (Experimentally Validated) &
  Make a sensor ignore legitimate inputs by using an additional infrared source to saturate the   sensors &
  Inject up to 3.33 times the   intended amount of fluid or 0.65 times for a 10-minute period &
  Low – Saturating sensor from   an external source \\ \hline
Reverse engineering exploitation &
  Marin et al \cite{41}   (03/2016) &
  Literature (Experimentally Validated) &
  Fully reverse-engineered the wireless communication protocol between all the peripherals of the insulin pump system &
  Compromise both the security and privacy of the patient &
  Passive attacks: Low - Radio interception and generation device \newline
  Active attacks: Medium - Reverse engineering\\ \hline
Reverse engineering exploitation &
  Li et al \cite{1} (07/2011) &
  Literature (Experimentally Validated) &
  No standard cryptographic mechanisms; the wireless protocol can be fully reverse-engineered, enabling eavesdropping, replay, and command manipulation attacks. &
  Compromise both the   security and privacy of the patient &
  Passive attacks: Low - Radio interception and generation device \newline
  Active attacks: Medium - Reverse engineering \\ \hline

\end{tabular}
% \end{adjustbox}

% \textcolor{red}{(Increase font size - consider landscape layout if necessary)}
\label{table3-2}
\end{table*}

%% file: Table_3-1.tex
% \usepackage{multirow}
\begin{table*}[!h]
\centering
\caption{Categorised attack vectors and common countermeasures in AID systems.}
% \begin{adjustbox}{width=1\textwidth}
\scriptsize
\begin{tabular}{|p{0.1\textwidth}|p{0.2\textwidth}|p{0.25\textwidth}|p{0.26\textwidth}|}
\hline
Security Aspect &
  Attack Vector &
  Description &
  Countermeasures \\ \hline
\multirow{2}{0.1\textwidth}{Confidentiality} &
  Eavesdropping \newline \cite{1}\cite{41} &
  Theft of information transmitted through a network &
  Cryptography; \newline
  Secure communication protocols;  \newline
  Proximity communication channels; \\ \cline{2-4} 
 &
  Do-it-yourself hacking\cite{15} &
  Patients extract data not provided by the products &
  Authentication \\ \hline
\multirow{6}{*}{Integrity} &
  Replay attacks\cite{73}\cite{7} &
  Intercept and retransmit valid data or commands &
   \\ \cline{2-3}
 &
  Bias injection attacks\cite{42} &
  A constant bias was injected into the sensor readings &
  \multirow{4}{0.45\textwidth}{
  Intrusion detection systems; \newline
  Control strategies assess; \newline
  Additional sensor for cross-check} \\ \cline{2-3}
 &
  False data injection attacks \cite{10}\cite{44} &
  Change transmitted data with random or synthesised data &
   \\ \cline{2-3}
 &

  AI/ML model attacks \cite{140&145}\cite{45} &
  Deviate the ML model’s   prediction during training or inference &
   \\ \cline{2-4} 
 &
  Computational attacks\cite{4} &
  Compromise the computational command or internal data &
  \multirow{2}{0.3\textwidth}{Formal specification} \\ \cline{2-3}
 &
  Malicious pump driver \cite{81} &
  Change pump I/O operations &
   \\ \hline
\multirow{5}{30pt}{Availability} &
  Denial-of-Service (DoS) /Distributed DoS \cite{81} &
  Flooding the target with   traffic results in the inability to respond timely &
  \multirow{5}{0.45\textwidth}{
    Secure communication protocols; \newline
    Redundant hardware or software components; \newline
    Formal specification; 
  } \\ \cline{2-3}
 &
  Ransomware Attacks \cite{48} &
  The malware encrypts the   system's data or locks users out, demanding a ransom to restore access &
   \\ \cline{2-3}
 &
  Jamming Attacks \cite{158} &
  Disrupting the wireless communication channels that the APS uses to operate &
   \\ \cline{2-3}
 &
  Firmware/ Software Corruption \cite{48} &
  Corrupting the firmware or   software that the APS relies on to function &
   \\ \cline{2-3}
 &
  Routing protocol information attacks \cite{99} &
  Attacks on routing   protocols to disrupt or manipulate the flow of information &
   \\ \hline
\end{tabular}
% \end{adjustbox}

\label{table3-1}
\end{table*}

%% file: Table_4-1.tex
% \usepackage{multirow}
\begin{table*}[!h]
\centering
\caption{Defence strategies proposed for AID systems.}
\begin{adjustbox}{width=0.9\textwidth}
\small
\begin{tabular}{|p{0.15\textwidth}|p{0.18\textwidth}|p{0.25\textwidth}|p{0.45\textwidth}|}
\hline
\multicolumn{2}{|p{0.25\textwidth}|}{Defence type} &
  Description &
  Methods \\ \hline
\multirow{4}{0.15\textwidth}{Protected communication} &
  Authentication &
  Verify the identity of users or devices &
  Fingerprint recognition\cite{52}\cite{55};
  Voiceprint analysis\cite{56}; 
  Physical Unclonable Functions\cite{68}\cite{70}; 
  Blockchain\cite{68}\cite{75} \\ 
 \cline{2-4} 
 &
  Cryptography &
  Transform   information into an unreadable format for unauthorized users &
  Rolling-code encryption\cite{1}; 
  AES-based encryption\cite{100}\cite{41}; 
  Asymmetric encryption \cite{52}; 
  Homomorphic encryption \cite{72} \\
 \cline{2-4} 
 &
  Secure communication protocols &
  Define rules for secure data transmission &
  RF-Harvested energy-based protocol\cite{48}; 
  Emergencies-resilient protocol\cite{77}; 
  Improved the Dexcom G4 protocol\cite{158}; 
  Mutual authentication  protocol\cite{155} \\ \cline{2-4} 
 &
  Proximity communication channels &
  Require close physical proximity between the communicating devices &
  Body-coupled communication\cite{1}; 
  Visible light communication\cite{53} \\
\hline
\multirow{3}{0.15\textwidth}{Intrusion detection systems} &
  Signature-based &
  Compare the real-time behaviour of the system against known security attacks &
  Define computational, data integrity, and communication attacks\cite{4};
  Define abnormal signal’s physical and content features\cite{6}; 
  Define replay and DoS attacks\cite{73}; 
  Infusion site failures \cite{149}; Pressure-induced sensor attenuations \cite{160} \\ \cline{2-4} 
 &
  Specification-based &
  Define  specifications of AID systems' normal operations and deem any violation as an intrusion &
  Formal specification \cite{159} with
  personalized thresholds\cite{5}\cite{81}; 
  statistical analysis\cite{109}; 
  declarative property \cite{4}; 
  outlier detection\cite{59}\cite{64}; 
  specification mining\cite{60}; 
  external and internal state\cite{61}\cite{62}; 
  insulin predictive model\cite{81}\\ \cline{2-4} 
 &
  Anomaly-based &
  Identify data patterns that have different data characteristics from normal instances &
  Model-based methods:\newline
  Physiological \cite{46}\cite{51}\cite{42}\cite{74}\cite{79}\cite{162}\cite{163}; \newline 
  Data-driven \cite{47}\cite{49}\cite{168}\cite{167}\cite{69}\cite{75}\cite{76}\cite{80}
  \cite{67}\cite{164}\cite{166}; \newline
  Model-free methods: \newline
  Supervised \cite{58}\cite{78};\newline
  Unsupervised \cite{9}\cite{54}\cite{57}\cite{65}\cite{161}; \newline
  Hybrid methods \cite{71}\cite{59}\cite{64}\cite{81} \\ \hline
\multicolumn{2}{|p{0.35\textwidth}|}{Control strategies assessment} &
  Compare the dosage with the identified patient’s insulin infusion pattern &
  Rule-based \cite{4}; 
  Data-driven\cite{3}\cite{10};
  Hybrid\cite{5} \\ \hline
\multicolumn{2}{|p{0.35\textwidth}|}{Redundant hardware or software components} &
  Use of multiple,   often independent, components that perform the same function &
  Hardware: 
  ECG monitor\cite{62}; Universal Software Radio Peripheral\cite{6}; Separate computation core\cite{48} \newline
  Software:  
  Replay attack detection filters\cite{7}; Auxiliary safety layer for controllers\cite{165}\\ \hline
\end{tabular}
\end{adjustbox}

\label{table4-1}
\end{table*}

%% file: Table_4-2.tex
\begin{table*}[!h]
\centering
\caption{Proposed cryptography-based approaches in AID systems.}
\begin{adjustbox}{width=1\textwidth}
\small
\begin{tabular}{|>{\raggedright\arraybackslash}p{0.15\textwidth}|>{\raggedright\arraybackslash}p{0.4\textwidth}|>{\raggedright\arraybackslash}p{0.1\textwidth}|>{\raggedright\arraybackslash}p{0.2\textwidth}|>{\raggedright\arraybackslash}p{0.2\textwidth}|}
\hline
    Encryption Method&Description&Key Sizes&Security Strength& Limitations\\
\hline
 Rolling code en/decoder \cite{1}& The remote controller encrypts an incrementing sequence counter as a shared key which is decrypted by the insulin pump to verify the counter within an acceptable range& 32–128 bits (Keeloq: 64-bit)& Provide basic security by preventing straightly intercepting the device series number and replay attacks&Not offer strong authentication. Not clarifying the management of the shared key.\\
 \hline
 AES-128 optimized for energy consumption \cite{41}& The remote control and the insulin pump share two independent symmetric cryptographic keys for encryption and authentication. Message size and transmission frequency are reduced to lower communication costs.& 128 bits& Strong security and relatively lower computational cost&Vulnerable if public keys are intercepted\\
 \hline
 ForkAE \cite{100}& Simultaneously performs encryption and authentication by creating two copies of ciphertext for the plaintext but adds an extra constant to one of the copies and encodes both blocks using the same encryption algorithm and key& 128 bits& Strong security and lower computational cost than AES-128&Vulnerable if the fork pattern is predictable or public keys are intercepted\\
 \hline
 RSA encryption using SHA generated 1024 key \cite{52}& An asymmetric cryptographic pattern is applied between a smartphone (communicate with sensors and actuators) and a control laptop& 1024 bits& Moderate security. Include data signing with the private key in every packet in case public keys are intercepted.&Higher computational cost. Prone to factorization and quantum attacks\\
 \hline
 A Fully Homomorphic Encryption scheme (CKKS) \cite{72}& A PID controller was employed to use an encrypted glucose signal as an input to produce encrypted control signals& Variable, typically in the range of hundreds of KB to MB& Very strong security and preserved privacy. &Computational cost is very high. Lack of division and comparison operations.\\
 \hline

\end{tabular}
\end{adjustbox}

\label{table4-2}
\end{table*}

%% file: Table_2-1.tex
% \usepackage{multirow}
\begin{table*}
\centering
\caption{Regulation standards and guidelines of cybersecurity in medical devices.}
\begin{adjustbox}{width=1\textwidth}
\small
\begin{tabular}{|p{0.1\textwidth}|p{0.3\textwidth}|p{0.4\textwidth}|p{0.4\textwidth}|}
\hline
Applied Area &
  Regulation Standards and Guidelines &
  Cybersecurity &
  Note \\ \hline
\multirow{4}{*}{US} &
  Policy for Device Software Functions and Mobile   Medical Applications &
  \multirow{2}{0.4\textwidth}{Require adequate cybersecurity to secure the correct functioning of medical devices} &
  \multirow{2}{0.4\textwidth}{Provide specific cybersecurity considerations} \\ \cline{2-2}
 &
  Section 3060 and 3060(a) of the 21st   Century Cure Acts &
   &
   \\ \cline{2-4} 
 &
  EO 14019 on America’s Supply Chains &
  Require detailed technical cybersecurity measures used by the product and related services &
  Not mandated for medical devices yet \\
 \cline{2-4}& NIST’s cybersecurity framework& Provides a catalogue and categorisation of typical cybersecurity activities and controls that should be considered by an enterprise& Encourage the private sector to determine its conformity needs, and then develop appropriate conformity assessment programs\\ \cline{2-4} 
 &
  FDA’s premarket and postmarket guidances for cybersecurity of medical devices&
  Provides a prescriptive and structured approach to cybersecurity risk management&
  Emphasise transparency in threat modelling and device labelling, but no conformity assessment program\\ \hline
\multirow{6}{*}{EU} &
  Medical Devices Regulation (MDR) &
  No provisions specifically labeled ‘cybersecurity’ &
  Highest priority \\ \cline{2-4} 
 &
  MDRAnnexex 1: General Safety and Performance Requirements (GSPR) &
  Require that medical devices do not fail due to any type of adversary attacking them &
  Imply minimum levels of encryption, access control, and physical cybersecurity \\ \cline{2-4} 
 &
  Network and Information Systems (NIS 2) Directive &
  Cybersecurity requirements of the general critical IT infrastructure &
  Integrated into Member States in various ways \\ \cline{2-4} 
 &
  Cyber Resilience Act and the Cyber Solidarity Act &
  Regulate the cybersecurity of any network-enabled devices &
  Supposed to entail more resilience (greater recovery from failure) than MDR \\ \cline{2-4} 
 &
  Security   standards: ISO 14155:2011 &
  Mostly developed by private actors and does not necessarily reflect best practices. &
  Allow usage of the standard for a fee, which becomes a barrier to adoption \\ \cline{2-4} 
 &
  General Data Protection Regulation (GDPR) &
  Protecting the fundamental right to privacy &
  Only considering cybersecurity requirements related to personal data \\ 
 \cline{2-4}& MDCG 201916: Guidance on Cybersecurity for medical devices& Offering recommendations to manufacturers, integrators, and operators on how to satisfy the cybersecurity requirements of the MDR&Provides broader, principle-based guidance on integrating cybersecurity into safety risk analysis\\
 \hline
\end{tabular}
\end{adjustbox}

% \label{table1}
\end{table*}

%% file: Table_2-3.tex
\begin{table*}
\centering
\setlength{\tabcolsep}{3pt}
\caption{Cybersecurity in representative AID system products.}
\begin{adjustbox}{width=1\textwidth}
\small
\begin{tabular}{|p{0.2\textwidth}|p{0.3\textwidth}|p{0.3\textwidth}|p{0.35\textwidth}|}
\hline
Product &
  Components &
  Main Functions &
  Cybersecurity Measures \\ \hline
CamAPS FX &
  User's smartphone (controller); \newline
  DANA/Ypsomed insulin pump; \newline
  Dexcom/Libre 3 CGM &
  Automated basal insulin delivery; \newline
  Manual bolusing for meals &
   Employs encryption, pseudonymisation, and anonymisation;  \cite{CamDiab-Privacy-2024} \\ 
   \hline
Medtronic 780G &
  MiniMed Insulin pump (controller integrated); \newline
  MiniMed CGM &
  Automated basal insulin delivery; \newline
  Automatic correction bolus delivery &
  No detail disclosure; aims to enable the highest levels of security and usability \cite{Medtronic-ProductSecurity-2022} \\ 
  \hline
Control IQ &
  User's smartphone (controller); \newline
  Tandem t:slim insulin pump; \newline
  Dexcom G6/G7 CGM; &
  Automated basal insulin delivery; \newline
  Automatic correction bolus delivery &
  Employs device authentication, message encryption, and message validation \cite{TandemControlIQRisks} \\ \hline
Omnipod 5 &
  User's smartphone (controller); \newline
  Omnipod insulin pumps; \newline
  Dexcom G6 CGM &
  Automatic basal insulin adjustments; \newline
  Manual bolusing for meals&
  Employs authentication, encrypted communication, and a verification system \cite{Insulet-ProductSecurity-2025}  \\ 
  \hline
DBLG1 System &
  DBLG1 controller; \newline
  Kaleido pump; \newline
  Dexcom G6 CGM &
  Automated basal insulin delivery; \newline
  Automatic correction bolus delivery &
  Employs data encryption, secured Bluetooth protocol and dedicated handset \cite{Diabeloop-DBLG1-DH22}  \\ 
  \hline
\end{tabular}
\end{adjustbox}

\label{table3}
\end{table*}

%% file: Table_4-3.tex
% \begin{sidewaystable}[!htbp]

% \begin{landscape}

% \begin{adjustbox}{width=1.0\textwidth}
{
\footnotesize  
\centering
\onecolumn
\begin{longtable}
% {|p{0.11\textwidth}|p{0.11\textwidth}|p{0.25\textwidth}|p{0.18\textwidth}|p{0.20\textwidth}|p{0.17\textwidth}|}
{|p{0.10\textwidth}|p{0.09\textwidth}|p{0.25\textwidth}|p{0.15\textwidth}|p{0.15\textwidth}|p{0.20\textwidth}|}
    \caption{Intrusion detection systems developed for AID systems.} \label{table4-3} \\

    \hline
        \textbf{Reference}&\textbf{Category} &\textbf{Detection method}&\textbf{Threats Evaluated}& \textbf{Dataset}&\textbf{Performance metric}\\
    \hline
    \endfirsthead

    \hline
        \textbf{Reference}&\textbf{Category} &\textbf{Detection method}&\textbf{Threats Evaluated}& \textbf{Dataset}&\textbf{Performance metric}\\
    \hline
    \endhead

    \multicolumn{6}{r}{Continued on next page} \\

    \endfoot

    \endlastfoot

     Zhang et al. (2013) \cite{6}& Signature-based& Monitor all the RF wireless communication. Identify and jam potentially malicious transactions& Unauthorized remote control
    & Simulated with glucose sensor and insulin delivery system &Detection rate: 100\%;FNR: 0\%
    Detection time: about 2.5 ms 
    \\\hline
    
     Khan et al. (2020) \cite{4}& Signature-based& Specify each attack as a one big step relation between the initial and final state  & Controller command manipulation attacks& Software simulation on a 2.6-GHz Intel Core i7 processor &CPU Time: 7.50 ± 0.3 * 10-4 s; Run time: 7.60 ± 0.3* 10-4 s\\\hline
     
     Walenda et al. (2023) \cite{73}& Signature-based& For replay attacks, detect consecutive repeated values. For DoS attacks, detect consecutive 0.& Replay attacks; DoS attacks
    & T1D patient simulator - GlucoSym Simulator&Qualitative illustration of detection \\
    \hline
    
     Howsmon et al. (2018) \cite{149}& Specification-based& Patient-specific metrics are calculated and compared to sliding window averages. Alarm when thresholds are exceeded.& Infusion site failure& 19 participants, 2 weeks&Sensitivity: 88.0\% (n = 25) FPR: 0.22 per day\\
     \hline
     
     Alshalalfah et al. (2019) \cite{109}& Specification-based& Combine formal methods with statistical analysis to evaluate the safety of control algorithms & Replay attacks
    & T1D patient simulator - UVA/PADOVA 2013 (5 patients, 24 hours)&Qualitative evaluation\\\hline
    
     Aliabadi et al. (2021) \cite{60}& Specification-based& Dynamically mining specifications based on their contribution to detecting every arbitrary attack& CGM spoofing attack; Stop basal injection; Resume basal injection& OpenAPS simulator&FPR: 13.5\%; FNR: 2\%
    \\\hline
    
     Astillo et al. (2021) \cite{59} & Specification-based& Incorporated specification with an outlier detection method & Manipulate random points of the glucose data within 1\% to 10\% deviation  & UVa/Padova simulator (2008) + Raspberry-Pis &AUROC: Attack rate 50\%-90\%: 99.94\% ± 0.04\%; Attack rate 10\%-30\%: 54.09\%± 2.5\% \\\hline
     
     Astillo et al. (2021) \cite{64}& Specification and anomaly-based& Extend SMDAps with timing-based malicious transmission detection and trustworthiness evaluation& Manipulate random points of the glucose data within 0\% to 50\% deviation & UvA/Padova simulator (2008) + Raspberry-Pis (2 virtual patients, 24 hours)&Accuracy under the hidden attack mode: 99.1\% ± 0.7\%
    \\\hline
    
     Prematilake et al. (2021) \cite{61}& Specification-based& Combined state transition rules, I/O access rules, and physiological rules to monitor both extrinsic state and internal state& Recurring bolus dose; Bolus unmatched with meal taking; Buffer overflow& Simulated with a safety-enhanced controller board and glucose-insulin model&Delay time: 253 ms; Energy overhead: 3\% (a 2500 mAh AA-sized battery)
    \\\hline
    
     Zhou et al. (2021) \cite{5}& Specification-based& Formal specified unsafe system context (physiological) and personalized thresholds using reinforcement learning& Availability attack; DoS attack; Integrity attack/memory fault & Simulated with Glucosym and UVA-Padova Simulator 2013 &F1-Score: 0.975 ± 0.005\\\hline
     
     Venugopalan et al. (2024) \cite{81}& Specification and anomaly-based& Combined formal methods (prove correct the most critical parts) and dictated safe bounds of insulin amounts to enable DL/ML to control the insulin pump safely.& Malicious ML; Malicious pump drivers & Real-world (1 individual, 6 months); UvA/Padova simulator (2008) (21 virtual humans) &Time In Range (TIR): Adults:92.88\% Adolescents: 82.38\%; Children: 85.42\%\\\hline
     
     Vega-Hernandez et al. (2009) \cite{46}& Anomaly-based (physiological model)& Incorporated the effect of the interstitial glucose measurements and the meal intake dynamics into the glucose-insulin metabolism model& Over- and sub-dosing& MATLAB SIMULINK simulation, 40 days with a normal patient routine&Qualitative evaluation\\\hline
     
     Facchinetti et al. (2013) \cite{47}& Anomaly-based (Data-driven model)& An individualized state-space model and a Kalman predictor & 1) spike on CGM 2) a loss of sensitivity of the CGM sensor 3) stop insulin delivery & UVA/Padova 2008, 100 simulated patients; 3 real T1DM subjects&CGM failure: FNs: < 5\%, FPs: < 10\%, CSII pump failure: Detection rate: 75\% within average 60 min, FPs: 10\%\\\hline
     
     Favero et al. (2014) \cite{49}& Anomaly-based (Data-driven model)& Extend \cite{47} for whole-day use by taking meals into account& 1) spike on CGM 2) consecutive large errors 3) meal*(1+e\%) 4) bias meal bolus 5) basal faults & UVA/Padova 2008, 100 simulated patients& Sensitivity: CGM faults:  ~ 85\% Meal faults: ~ 90\% Bolus faults: ~ 90\% Basal failure: 50\% \\\hline

       Meneghetti et al. (2020) \cite{168} & Anomaly-based (Data-driven model) & Distinguish insulin pump faults and missed meal announcements by introducing 2 sets of parameters of model  & Night-time pump occlusion ; missed meal announcement ; meal-time pump occlusion  & UVA/Padova Type 1 Diabetic Simulator 2013 & Pump faults: sensitivity: ~81.3\% FPR: ~0.15/day  missed meals: sensitivity: ~86.8\% FPR: ~0.15/day\\\hline

      Meneghetti et al. (2023) \cite{167} & Anomaly-based (Data-driven model) & Extend \cite{168} with analyse various statistical properties  & Insulin suspension; CGM missed samples and pressure-induced sensitivity losses; missed meal announcements  & UVA/Padova Type 1 Diabetic Simulator 2013 & Recall: 91.3\% FP: 1 every 10 days\\\hline

     Tosun et al. (2022) \cite{66}& Anomaly-based (physiological model)& Kalman filter and a $\chi^2$ test. Identify meal disturbance by an online meal estimator and a time-varying threshold& Bias injection attack on glucose sensor& Medtronic Virtual Patient (MVP) model (1 virtual subject) + PID controller&Qualitative evaluation\\\hline
     
    Tosun and Teixeira (2023) \cite{74}& Anomaly-based (physiological model)& Kalman filtering and sequential change detection (detecting an abrupt change)& Deterministic sensor deception attacks (a slowly increasing bias injection attack& Bequette model (glucose)&Qualitative evaluation (the $\chi^2$ test maximizes the best-case detection delay, while the CUSUM test minimizes the worst-case delay)\\\hline
     
    Tosun et al. (2024) \cite{79}& Anomaly-based (physiological model)& The prediction error of a time-varying Kalman filter is statistically evaluated to detect anomalies under meal disturbances with known time& Sensor bias injection attack& Medtronic Virtual Patient (MVP) model (1 virtual subject) + PID controller&Qualitative evaluation \\\hline

     Turksoy et al. (2016) \cite{164}& Anomaly-based (Data-driven model)& Non-linear first principle dynamic model with an unscented Kalmen filter. PCA and K-means clustering hired for fault detection & Faulty CGM measurements & Data from 51 subjects & Sensitivity: 84.2 \%; Average detection time: 2.8 min\\\hline

     Olney et al. (2022) \cite{69}& Anomaly-based (Data-driven model)& Non-linear autoregressive neural networks& Incorrect sampling of the target biosignal (sensor data)& D1NAMO CGM dataset (9 human subjects) + FPGA&Average RMSE: Software: 0.428 Hardware: 0.455\\\hline
     
     Maity et al. (2023) \cite{76}& Anomaly-based   (Data-driven model)& Learn surrogate model for safety-certified actions using RNN, then periodically relearn model coefficients to check deviation& Insulin cartridge problem& UVA/Padova 2018&Difference in model coefficient values: \( > \) 0.27 \\\hline
     
     Rahmadika et al. (2023) \cite{75}& Anomaly-based   (Data-driven model)& BiLSTM and augments the security through integrating blockchain technology& Arbitrarily changing a feature value (increase or decrease 10\% - 50\%)& UVA/Padova 2008&Average RMSE: 1.3482 (5 timesteps), 1.9959 (7 timesteps), 2.6099 (9 timesteps)\\\hline
     
     Zhou et al. (2022) \cite{67}& Anomaly-based   (Data-driven model)& Extract the specification of unsafe control actions and add them as regularization terms in a semantic loss function& Designed small but malicious input perturbations for both sensor and  commands & Glucosym simulator; UVA-Padova Simulator 2013 &Robustness error: Reduce on average up to 54.2\%\\\hline
     
      Meneghetti et al. (2018) \cite{161}& Anomaly-based   (Model free)& Local outlier factor and connectivity-based outlier factor AD methods & insulin pump failure(suspensions occurring at night for 6 hours) & UVA/Padova Type 1 Diabetic Simulator 2013 & Precision: ~75\%; Recall ~60\%\\\hline

     Meneghetti et al. (2019) \cite{54}& Anomaly-based   (Model free)& Unsupervised anomaly detection with optimum feature set selection& Insulin pump faults& Padova/UVA T1D simulator 2018, 100 adults, 30 days&Histogram-based outline score: Sensitivity: 0.87  FPR: 0.08/day Isolation forest:  Sensitivity: 0.85  FPR: 0.06/day\\\hline

     Meneghetti et al. (2022) \cite{9}& Anomaly-based   (Model free)& Unsupervised anomaly detection& Infusion site failure& Clinical dataset (N=20)&Sensitivity: 0.75 (15 out of 20); FPR: 0.08/day\\\hline

     Meneghetti et al. (2020) [57]& Anomaly-based   (Model free)& Unsupervised multi-dimensional data-driven anomaly detection. Cut several start and end points to enlarge the difference between normal and abnormality& Suspension of insulin delivery& UVA/Padova T1D Simulator 2013; 7 patients, 1 month in free-living conditions&Recall: ~80\% Precision: \( > \)90\%\\\hline
     
     Herrero et al. (2022) \cite{65}& Anomaly-based (Model-free)& Detect whether CGM data belong to the individual using SVM binary classifier with 12 standard glycemic metrics at 24h, day, night, breakfast, lunch, and dinner& Reidentification& REPLACE-BG (NCT02258373), 226 adult, 6 month&Accuracy: 86.8\% ± 0.16; F1-score: 0.86 ± 0.16 (window length: 15 days)\\\hline
     
     Astillo et al. (2022) [71]& Anomaly-based (Data-driven  and Model-free)& Deep learning-based anomaly detection system with federated learning& False sensor data& UVa/Padova simulator (2008)&Accuracy: 99.17\%; F1-score: 99.16\%\\\hline

      Judith et al. (2023) \cite{78}& Anomaly-based   (Model free)& PCA for feature reduction and multi-layer perceptron (MLP) for classification& Man-in-the-middle attacks& WUSTL-EHMS-2020 dataset &Sensitivity: 95.4\% Specificity: 100\%\\\hline

\end{longtable}
\twocolumn
}
% \end{adjustbox}
% \end{landscape}

% \end{sidewaystable}

%% file: Table_5-1.tex
% \usepackage{multirow}
% \begin{table*}[]
% \centering
% \caption{Available Resources and Their Utilization in Research}
% \begin{tabular}{|L{30pt}|L{65pt}|L{150pt}|L{30pt}|L{30pt}|L{80pt}|}
% \usepackage{multirow}
% \usepackage{longtable}
% Note: It may be necessary to compile the document several times to get a multi-page table to line up properly
% \usepackage{multirow}
\begin{table*}[!htbp]
\centering
\caption{Available resources and utilisation in AID cybersecurity research.}
\begin{adjustbox}{width=1.0\textwidth}
\scriptsize 
\begin{tabular}{|p{0.06\textwidth}|p{0.12\textwidth}|p{0.28\textwidth}|p{0.12\textwidth}|p{0.08\textwidth}|p{0.24\textwidth}|}
\hline
Type &
  Resource Name &
  Description &
  Focus area &
  Availability &
  Usage in Research \\ \hline
\multirow{4}{30pt}{Hardware simulation} &
  Commercial diabetes medical products &
  Widely used hybrid closed-loop insulin delivery system or its components&
  Device functionality &
  off-the-shelf &
 Identify cyber vulnerabilities in APS \cite{1}\cite{41}; \newline
 Evaluate cryptography and authentication \cite{52} \\ \cline{2-6} 
 &
  Universal Software Radio Peripheral (USRP) &
  A software radio board that can intercept radio communications and generate wireless signals &
  Wireless communication &
  off-the-shelf &
  Hijack APS \cite{1}\cite{41}; \newline
  Monitor wireless transactions\cite{6} \\ \cline{2-6} 
 &
  Biostator II\cite{105}&
  An ultra-low-power ASIC implementation of an AP controller &
  Implant functionality &
  Open source &
  Evaluate the energy efficiency of security communication protocol\cite{48} \\ \cline{2-6} 
 &
  Microcontroller board &
  Simulate controller algorithms on board (e.g., Arduino Uno-R3\cite{53}\cite{169}\cite{170}; EFM32WG\cite{61}, Arduino Nano 33 kit , nRF52840 development kit, and Galaxy S20 with Android 13\cite{77}; ESP8266 development kit \cite{127}&
  Integrated function &
  off-the-shelf &
  Incorporated visible light access control channel \cite{53}; \newline
  Implement rule check mechanisms \cite{61} \\ \hline
\multirow{4}{40pt}{T1D patient simulator} &
  UVA/Padova 2008 \cite{112} &
  An FDA-approved nonlinear simulator captures interpatient variability \cite{111} &
  In silico trials &
  Open source &
  Simulate attacked monitoring data and evaluate intrusion detection systems\cite{47}\cite{49} \cite{59}\cite{71}\cite{81} \\ \cline{2-6} 
 &
  UVA/Padova 2013 \cite{34} &
  Changes patient’s physiology over time &
  In silico trials &
  Licensed &
  Evaluated control algorithms\cite{109}; \newline
  Unsupervised anomaly detection \cite{57}; \newline
  Homomorphic encryption\cite{72}; \newline
  Supervised anomaly detection\cite{127} \\ \cline{2-6} 
 &
  UVA/Padova 2018 \cite{110} &
  Introduce intraday variability of insulin sensitivity, time-varying distributions of the patients’ therapy parameters, and a “dawn” phenomenon model &
  In silico trials &
  Not public open &
  Feature selection and evaluate unsupervised anomaly detection algorithms\cite{8}; Early detection of unknown-unknowns attacks \cite{76} \\ \cline{2-6} 
 &
  Glucosym \cite{35} &
  Online simulator that allows different control algorithms to be integrated to test their efficiency with 10 real-world patient profiles &
  In silico trials &
  Open source &
  Simulate attack and evaluate defence methods \cite{73} \\ \hline
\multirow{5}{30pt}{Pharmaco-kinetic model} &
  Hybrid physiologically-based pharmacokinetic model\cite{106} &
  Simulate the physiological response of a T1D patient considering recirculation dynamics &
  Closed-loop insulin delivery algorithms &
  Equations and parameters available &
  Evaluate the overhead of runtime as an Android service \cite{107} \\ \cline{2-6} 
 &
  Medtronic Virtual Patient (MVP) model\cite{113} &
  Explicitly describe intraday variation in metabolic parameters &
  Closed-loop insulin delivery algorithms &
  Equations available &
  Compare the performance of nonlinear filters in online drift detection of CGM\cite{51}\cite{66}\cite{42} \\ \cline{2-6} 
 &
  Ackerman’s Linear model \cite{121} &
  Simulate glucose regulation system which is an abstraction of reality &
  Simulate blood glucose control &
  Equations available \cite{123} &
  Simulate and detect replay attack\cite{7} \\ \cline{2-6} 
 &
  Bergman Minimal Model \cite{123} &
  Describes the dynamics of the glucose uptake after an external stimulus in humans with a simplistic structure &
  Simulate blood glucose control &
  Equations available &
  Evaluate homomorphic encryption \cite{72} \\ \cline{2-6} 
 &
  Other Mathematical models for T1D \cite{83} &
  Providing a comprehensive overview of the modelling of the biological processes of APS &
   & 
   &
   \\ \hline
\multirow{2}{30pt}{Controller algorithm} &
  OpenAPS \cite{117} &
  Implements a controller component of APS in JavaScript &
  Controller’s function &
  Open source &
  Evaluate specification mining technique \cite{60} \\ \cline{2-6} 
 &
  Other control strategies \cite{83} &
  Providing a comprehensive overview of available control strategies of APS &
   &
   &
   \\ \hline
\multirow{2}{30pt}{Closed-loop testbed} &
  Closed-loop APS testbed \cite{33} &
  Made up of two APS simulators, UVA/Padova and Glucosym, and a fault injection engine &
  APS simulation and faults   data generation &
  Open-source &
  Evaluate runtime monitoring unsafe control commands \cite{67}\cite{119} \\ \cline{2-6} 
 &
  MATLAB Simulink model \cite{122} &
  Simulate a CGMS infusion pump for blood glucose regulation &
  APS simulation &
  Open-source &
  Design attacks to compromise sensors or controllers \cite{2} \\
 \cline{2-6}& Hardware-in-the-Loop Simulation \cite{169}& Simulation of AP using the Bergman Minimal Model (BMM)& APS simulation& Open-source&Simulated the responses of template attack and DoS attack\\
 \cline{2-6}& Hardware-in-the-Loop Simulation \cite{170}& An low-cost HIL platform for testing control strategies for AP& APS simulation& Open-source&Simulated the PID and linear quadratic regulator (LQR) controllers\\ 
  \cline{2-6}&  LoopInsightT1 \cite{LoopInsightT1} & An in-browser simulation platform for modelling closed-loop interactions between individuals with T1D and AID systems& APS simulation& Open-source&Integrated multiple physiological models, such as the UVA/Padova and the Cambridge model, with several control algorithms, and offers interactive visualisations\\ \hline

\multirow{7}{30pt}{Datasets} &
  OpenHumans \cite{39} &
  Collected by \#OpenAPS project, donated by its users &
  BG reading, device function, patient activity &
  Apply for access &
   \\ \cline{2-6} 
 &
  JCHR - DCLP3 \cite{38} &
  An international diabetes   closed-loop trial dataset recorded 6-months from clinical trials with 168   patients &
  BG readings, insulin doses, and other relevant data &
  Publicly-available &
  Test the simulation performance of APS testbed \cite{33} \\ \cline{2-6} 
 &
  D1NAMO \cite{115} &
  20 healthy subjects and 9 patients with type-1 diabetes in real-life conditions. &
  ECG, breathing, glucose level and annotated food pictures &
  Publicly-available &
  Evaluating glucose prediction model \cite{114}\cite{69} \\ \cline{2-6} 
 &
  DICARDIA diabetic dataset\cite{116} &
  A diabetic patient database to study cardiovascular autonomic neuropathy (CAN) &
  stress ECG is presented &
  Publicly-available &
  Study the correlation of ECG intervals with blood glucose levels \cite{62} \\ \cline{2-6} 
 &
  REPLACE-BG dataset\cite{118} &
  CGM data and self-monitoring blood glucose of 226 adult participants with T1D for 26 weeks &
  BG data &
  Publicly-available &
  Re-identify CGM data\cite{65} \\ \cline{2-6} 
 &
  OhioT1DM Dataset\cite{126} &
  8 weeks’ worth of continuous glucose monitoring, insulin, physiological sensor, and self-reported life event data for each of 12 people with type 1 diabetes &
  BG level prediction &
  Publicly-available &
  Demonstrate the impact of adversarial inputs on the predictions of the targeted ML model \cite{45} \\ \cline{2-6} 
 &
  WUSTL-EHMS-2020 dataset \cite{124} &
  Collects both the network flow metrics and patients' biometrics, labelled as normal and attack instance &
  Network flow, biometrics &
  Publicly-available \cite{125} &
  Detect man-in-the-middle attacks \cite{78} \\ \hline
\end{tabular}
\end{adjustbox}
\label{table5-1}
\end{table*}